\documentclass[12pt,onecolumn,oneside,letterpaper,peerreview]{IEEEtran}

\usepackage{amsmath}
\usepackage{cite}
\usepackage{graphicx}
\usepackage{float}
\usepackage{times}
\usepackage{latexsym}
\usepackage{bm}
\usepackage{amssymb}
\usepackage[center]{caption2}
\usepackage{stfloats}
\usepackage{array}
\usepackage{fancyhdr}
\usepackage{cite,graphicx,amssymb}
\usepackage{psfrag}
\usepackage{multirow}

\usepackage{color}

\ifCLASSINFOpdf

\else

\fi

    \def\Complex{{\rm\rule[.23ex]{.03em}{1.1ex}\kern-.3em{C}}}

    \newcommand{\be}{\begin{equation}} \newcommand{\ee}{\end{equation}}
    \newcommand{\bea}{\begin{eqnarray}} \newcommand{\eea}{\end{eqnarray}}
    \newcommand{\benum}{\begin{enumerate}} \newcommand{\eenum}{\end{enumerate}}

    \newcommand{\qd}{{\bf d}}

    \newcommand{\qh}{{\bf h}}

    \newcommand{\qn}{{\bf n}}

    \newcommand{\qu}{{\bf u}}

    \newcommand{\qx}{{\bf x}}
    \newcommand{\qy}{{\bf y}}
    \newcommand{\qz}{{\bf z}}

    \newcommand{\qA}{{\bf A}}
    \newcommand{\qB}{{\bf B}}

    \newcommand{\qG}{{\bf G}}
    \newcommand{\qH}{{\bf H}}
    \newcommand{\qI}{{\bf I}}

    \newcommand{\qQ}{{\bf Q}}
    \newcommand{\qR}{{\bf R}}
    
    \newcommand{\qT}{{\bf T}}
    \newcommand{\qU}{{\bf U}}
    \newcommand{\qV}{{\bf V}}
    \newcommand{\qW}{{\bf W}}
    \newcommand{\qX}{{\bf X}}

    \newcommand{\qXi}{{\boldsymbol \Xi}}
    
    \newcommand{\qLambda}{{\boldsymbol \Lambda}}

    \newcommand{\qOmega}{{\boldsymbol \Omega}}
    \newcommand{\qPi}{{\boldsymbol \Pi}}

    \newcommand{\bbR}{{\mathbb R}}
    \newcommand{\bbC}{{\mathbb C}}

    \newcommand{\calF}{{\mathcal F}}
    
    \newcommand{\calJ}{{\mathcal J}}

    \newcommand{\calS}{{\mathcal S}}
    
    \newcommand{\calV}{{\mathcal V}}

    \newcommand{\diag}{{\sf diag}}
    
    \newcommand{\tr}{{\sf tr}}

    \newcommand{\bl}[1]{\color{black}#1}

\newtheorem{prop}{Proposition}
\newtheorem{alg}{Algorithm}

\newtheorem{example}{Example}

\makeatother

\begin{document}
\title{Low-Complexity MIMO Precoding for Finite-Alphabet Signals}

\author{\IEEEauthorblockN{Yongpeng Wu, Chao-Kai Wen, Derrick Wing Kwan Ng, \\
Robert Schober, and Angel Lozano}

\thanks{This paper was presented in part at IEEE ICC 2016.}

\thanks{Y. Wu is with Institute for Communications Engineering,  Technical University of Munich,
Theresienstrasse 90, D-80333 Munich, Germany (Email:yongpeng.wu2016@gmail.com).
The work of Y. Wu is supported by TUM University Foundation Fellowship.}

\thanks{D. W. K. Ng is with the School of Electrical Engineering and Telecommunications, University of New South Wales, Sydney, N.S.W.,
Australia (E-mail: w.k.ng@unsw.edu.au).}

\thanks{C. K. Wen is with the Institute of Communications Engineering, National Sun Yat-sen University, Kaohsiung 804,
Taiwan (Email:   chaokai.wen@mail.nsysu.edu.tw).}

\thanks{R. Schober is with the Institute for Digital Communications, Universit\"{a}t Erlangen-N\"{u}rnberg,
Cauerstrasse 7, D-91058 Erlangen, Germany (Email: robert.schober@fau.de).}

\thanks{A. Lozano is with Universitat Pompeu Fabra, 08018, Barcelona, Spain (Email: angel.lozano@upf.edu). His work is
supported by Project TEC2015-66228-P (MINECO/FEDER, UE) and by the European Research Council under the H2020 Framework Programme/ERC grant agreement 694974.
}
}

\maketitle

\begin{abstract}
This paper investigates the design of precoders for single-user multiple-input multiple-output (MIMO) channels, and in particular for finite-alphabet signals.
Based on an asymptotic expression for the mutual information of channels exhibiting line-of-sight components and rather general antenna correlations,
{\bl precoding structures that decompose the general channel into a set of parallel subchannel pairs are proposed.
Then, a low-complexity iterative algorithm is devised to maximize the sum mutual information
of all pairs. The proposed algorithm  significantly
reduces the computational load of existing approaches with only minimal loss in performance.}
The complexity savings increase with the number of transmit antennas and with the cardinality of the signal alphabet, making it possible to support values thereof that were {\bl unmanageable} with existing solutions.
Most importantly, the proposed solution does not require instantaneous channel state information (CSI) at the transmitter, but only statistical CSI.
\end{abstract}

\newpage

\section{Introduction}

%The topic of design of precoders in a multiple-input multiple-output MIMO channel has gathered momentum in recent years.
Although complex Gaussian signals are capacity-achieving under perfect channel state information (CSI) at the receiver, signals conforming to discrete constellations are transmitted in practice.
For such signals, the capacity-achieving approach---allocating more power to stronger channels---can be quite suboptimal, as illustrated for parallel channels in \cite{Lozano2006TIT} and \cite{Lozano2008TCOM}, and hence it is of interest to devise suitable precoders.

For multiple-input multiple-output (MIMO) channels with instantaneous CSI at the transmitter, an optimal linear precoder design was put forth in \cite{Xiao2011TSP}, building upon earlier works
\cite{Xiao2008GlobalCom,Xiao2009ICC,Payaro2009,Lamarca2009,Perez-Cruz2010TIT}.
%In the mean time, when perfect instantaneous CSI is available at the transmitter, the MIMO channel can be decomposed into a set of parallel subchannels.
In turn, \cite{Mohammed2011TIT} proposed to group MIMO subchannels in pairs and design the relevant parameters within each pair and among pairs to increase the mutual information with finite-alphabet signals. This significantly reduces
 the complexity of the precoder optimization with little loss in mutual information.
Recently, this idea was extended to pair multiple subchannels on the basis of a per-group precoding (PGP) technique \cite{Ketseoglou2015TWC,Ketseoglou2016}.
As an alternative way of reducing the computational load, a precoder design that optimizes a lower
bound of the mutual information (rather than the actual mutual information) was set forth in \cite{Zeng2012WCL}.

With only statistical CSI available at the transmitter, the MIMO precoding design
for capacity-achieving Gaussian signals was addressed in \cite{tulino2005TIT,tulino2006capacity,Gao,JWang2012TSP,Zhang2013JSAC,Wu2014TSP,Wu2016TIT}.
For discrete signals, an iterative precoding algorithm was proposed in \cite{zeng2012linear} for the Kronecker channel model, yet the complexity of this complete-search algorithm is exponential in the number of transmit antennas and,
even with modest numbers thereof (say, eight), it becomes {\bl unmanageable}.

The premise of instantaneous CSI at the transmitter is reasonable when users are static or slowly moving, such that the fading remains constant for a sufficiently long time.
%Then, the CSI can either be estimated accurately at the receiver via uplink training and then sent to the transmitters through dedicated feedback links for frequency division duplex systems, or obtained by exploiting the reciprocity of uplink and downlink channel for time division duplex systems.
With fast moving users,
%the channel varies rapidly so that the coherence time of the channel is much shorter than the round-trip delays introduced for collecting the CSI. In this case, the obtained instantaneous CSI at the transmitter is completely outdated. For these scenarios, the obtained instantaneous CSI is useless and we have to design the precoder based on
a more appropriate premise is to consider only statistical CSI at the transmitter.
%However, the statistical CSI loses many useful information of the channel such
%as the phase position, which may be important,  especially for future
%communication systems employing a large number of antennas and high order modulation \cite{Krishnan2015TSP,Krishnan2015TVT}.
%Therefore, for some scenarios where the mismatch
%between the obtained CSI and the instantaneous channel
%is not large, this imperfect CSI is still used for the precoder design.
%
%In practice, the above-mentioned three types of users may exist
%simultaneously in a cell. Therefore, it is necessary to find a general
%precoder design which applies to all these users in practice communication systems.
%Another important issue is that the implementation
%burden of this precoder design should be affordable.
%
This paper proposes a low-complexity precoder design for rather general single-user MIMO channels with finite-alphabet inputs and statistical CSI at the transmitter.
The contributions of the paper are as follows.

\begin{enumerate}
  \item An asymptotic (in the number of antennas) expression is derived for the mutual information  of MIMO channels {\bl with finite alphabet inputs} and    correlated Rice fading.
  %The obtained expression takes into account the line of sight (LOS) effect and the mutual dependence of the transmit and receive ends analytically.

   \item By exploiting the spatial characteristics of the adopted MIMO channel model, the left singular matrix of the optimal precoder is obtained. For positive Rice factors, the result differs from the one obtained in \cite{zeng2012linear} for the Kronecker channel model.

  \item Structures are established for the power allocation matrix and the right singular matrix of the precoder.
  These structures decouple the data streams over parallel equivalent subchannels, eliminating the need for a complete search of the entire signal space during the precoder optimization. The complexity of such optimization is thereby reduced by an exponential order of magnitude.

  \item A novel low-complexity iterative algorithm is devised for the precoder optimization. This  algorithm drastically reduces the computational load, but with only minimal loss---established on the basis of the 3GPP spatial channel model (SCM) \cite{Salo2005}---in performance.

  \item  Some special cases are investigated, chiefly massive MIMO where the additional structure in the channel simplifies the derived algorithm.
 % revealing, e.g., massive MIMO that under statistical CSI the number of subchannels that can  be paired  may be larger than under instantaneous CSI \cite{Mohammed2011TIT,Ketseoglou2015TWC}.
  %This results in an unexpected performance gain with statistical CSI.
  %We provide Example \ref{exp:4_1} to elaborate  on this point in details.

\end{enumerate}

The remainder of the paper is organized as follows. Section II briefly introduces some notation and describes the channel model.
In Section III, the complete-search algorithm is reviewed and an idea proposed for reducing its computational complexity.
Building on this idea, Section IV proposes a low-complexity precoding approach. Numerical results are provided in Section V, and conclusions are drawn in Section VI.

\section{Preliminaries}

\subsection{Notation}
The following notation is adopted throughout:
%Column vectors are represented by lower-case bold-face letters,
%and matrices are represented by upper-case bold-face letters. Superscripts $(\cdot)^{T}$, $(\cdot)^{*}$, and $(\cdot)^{H}$
%stand for the matrix/vector transpose, conjugate, and conjugate-transpose operations, respectively.
%$\rm{det}(\cdot)$ and $\rm{tr}(\cdot)$  denote the matrix determinant and trace operations, respectively.
Superscripts $(\cdot)^{T}$, $(\cdot)^{*}$, and $(\cdot)^{H}$
stand for the matrix/vector transpose, conjugate, and conjugate-transpose operations, respectively,
$\diag\{\bf{b}\}$  denotes a diagonal matrix containing
the entries of vector $\bf{b}$,
 $\diag\{\bf{B}\}$  denotes a diagonal matrix containing in the main diagonal
the diagonal elements of matrix $\mathbf{B}$,
 $\odot$ and $\otimes$ denote the entry-wise
and the Kronecker product of two matrices, respectively, ${\rm vec} (\mathbf{A}) $ is a column vector
containing the stacked columns of matrix $\mathbf{A}$,
$[\mathbf{A}]_{mn}$ denotes the $(m,n)$th entry of matrix $\mathbf{A}$, $[\mathbf{a}]_{m}$ denotes the $m$th entry
of vector $\mathbf{a}$,  $\left\| \cdot \right\|_F$ denotes the Frobenius norm,
and $E\left[\cdot \right]$ represents the expectation with respect to the random variable inside $\left[\cdot \right]$,
which can be a scalar, vector, or matrix.
Finally, $D \mathbf{A}$ denotes the integral measure for the real and imaginary parts of the entries of $\mathbf{A}$. That is, for an $n \times m$ matrix $\mathbf{A}$,
\begin{equation}
D \mathbf{A} = \prod_{i=1}^n \prod_{j=1}^m
\frac{d {\rm{Re}}{[\mathbf{A}]_{ij}}  \, d {\rm{Im}}{[\mathbf{A}]_{ij}} }{\pi}
\end{equation}
where $\rm{Re}[\cdot]$ and $\rm{Im}[\cdot]$ return the real and imaginary parts, respectively.

\subsection{Channel Model}\label{sec:model}

Consider a single-user MIMO channel where a transmitter and a receiver
are equipped with $N_{\mathrm{t}}$ and $N_{\mathrm{r}}$ antennas, respectively. The received signal
$\qy\in{\mathbb C}^{N_{\mathrm{r}}}$ can be
written as
\begin{equation}\label{eq:x}
 \qy = \qH \qx + \qn
\end{equation}
where $\qH \in{\mathbb C}^{N_{\mathrm{r}} \times N_{\mathrm{t}}}$ is a random channel matrix whose $(i,j)$th entry denotes the complex fading coefficient between the $j$th transmit and the $i$th receive antenna\footnote{\bl The channel model in (\ref{eq:x})
is only intended for Example 1, to explain the basic idea behind the low complexity design; it does not represent the jointly correlated Rician fading model analyzed in Section IV.}, $\qx \in{\mathbb C}^{N_{\mathrm{t}}}$  denotes the
zero-mean transmitted vector with covariance matrix  $\boldsymbol{\Sigma}_{\qx}$, and
$\qn \in{\mathbb C}^{N_{\mathrm{r}}}$ is the zero-mean
complex Gaussian noise vector with covariance matrix $\qI_{N_{\mathrm{r}}}$.
The transmit vector $\qx$ satisfies the power constraint
\begin{equation}\label{x_constraint_2}
{{\rm{tr}}\big(\boldsymbol{\Sigma}_{\qx} \big)} \leq P.
\end{equation}%
Based on the available CSI, and subject to the power constraint, we want to optimize $\boldsymbol{\Sigma}_{\qx}$
to maximize the spectral efficiency.

%\section{Precoder Design}
%\label{sec:transmit_design}

\section{Precoder Designs for Single-User MIMO Channel}
\label{sec:transmit_design}

%In this section, we review the complete-search approach to the optimization of $\boldsymbol{\Sigma}_{\qx}$ and introduce an idea to reduce its computational load.

Let $\qx = {{{\bf{B}}} \,{{\bf{d}}}}$, where $\mathbf{B} \in \mathbb{C}^{N_{\mathrm{t}} \times N_{\mathrm{t}}}$ is the precoder
whereas ${{\bf{d}}} \in \mathbb{C}^{N_{\mathrm{t}} \times 1}$ is a signal vector whose entries are drawn independently from an equiprobable $M$-ary constellation; there are $M^{N_{\mathrm{t}}}$ possible signal vectors, the $m$th of which is denoted by ${\mathbf{d}}_{m}$.
The precoder admits the singular value decomposition (SVD)
$\mathbf{B} = \mathbf{U}_{{\mathrm{\bf{B}}}} \boldsymbol{\Lambda}_{{\mathrm{\bf{B}}}} \mathbf{V}_{{\mathrm{\bf{B}}}}$
where $\boldsymbol{\Lambda}_{{\mathrm{\bf{B}}}} \in \mathbb{C}^{N_{\mathrm{t}} \times N_{\mathrm{t}}}$
is diagonal while $\mathbf{U}_{{\mathrm{\bf{B}}}} \in \mathbb{C}^{N_{\mathrm{t}} \times N_{\mathrm{t}}}$ and $\mathbf{V}_{{\mathrm{\bf{B}}}} \in \mathbb{C}^{N_{\mathrm{t}} \times N_{\mathrm{t}}}$ are unitary.
While, with Gaussian signaling, ${{\bf{d}}}$ would be unitarily invariant and thus $\mathbf{V}_{{\mathrm{\bf{B}}}}$ would be {\bl an identity matrix}, for the signals at hand $\mathbf{V}_{{\mathrm{\bf{B}}}}$ plays an important role.

When Gaussian-signal precoding solutions are applied to discrete constellations, the performance suffers because,
in the face of major power discrepancies among MIMO subchannels, these solutions insist on beamforming over an extensive range of signal-to-noise ratios (SNRs), well beyond the point where beamforming is appropriate for a discrete constellation.
With beamforming, signalling occurs only over the dominant subchannel, which causes a performance loss with discrete signals \cite{Xiao2011TSP,zeng2012linear}.
By properly designing $\mathbf{U}_{{\mathrm{\bf{B}}}}$, $\boldsymbol{\Lambda}_{{\mathrm{\bf{B}}}}$, and $\mathbf{V}_{{\mathrm{\bf B}}}$, this loss can be eliminated \cite{Xiao2011TSP,zeng2012linear}.
The matrix $\mathbf{V}_{{\mathrm{\bf{B}}}}$ mixes the $N_{\mathrm{t}}$ original signals into $N_{\mathrm{t}}$ beams, then $\boldsymbol{\Lambda}_{{\mathrm{\bf{B}}}}$ allocates power to those beams, and finally $\mathbf{U}_{{\mathrm {\bf{B}}}}$ aligns them spatially as they are launched onto the channel.
With a proper choice of $\mathbf{V}_{{\mathrm{\bf{B}}}}$, in particular, all the $N_{\mathrm{t}}$ signals can be
effectively transmitted even if only a single beam is active.

%The next example illustrates the role of $\mathbf{U}_{{\mathrm{\bf{B}}}}$, $\boldsymbol{\Lambda}_{{\mathrm{\bf{B}}}}$, and $\mathbf{V}_{{\mathrm{\bf{B}}}}$.

\vspace{2mm}

\begin{example}
\label{4_4}

In a $4 \times 4$ MIMO channel,
\begin{equation}
\label{eq:y_4_exp}
{\bf{y}} = {\bf{H}} \, {{\bf{U}}_{{{\mathrm{\bf{B}}}}}}{{\bf{\Lambda }}_{{{\mathrm{\bf{B}}}}}}{{\bf{V}}_{{{\mathrm{\bf{B}}}}}} \, {\bf{d}} + {\bf{n}}
\end{equation}
where ${{\bf{H}}} = \mathbf{U}_{{\mathrm{\bf{H}}}} \boldsymbol{\Lambda}_{{\mathrm{\bf{H}}}} \mathbf{V}_{{\mathrm{\bf{H}}}}$
and ${\bf{d}} = [d_1,d_2,d_3,d_4]^T$.
{\bl Here, $\mathbf{U}_{{\mathrm{\bf{H}}}} \in \mathbb{C}^{4 \times 4}$ and $\mathbf{V}_{{\mathrm{\bf{H}}}} \in \mathbb{C}^{4 \times 4}$
are unitary matrices, and $\boldsymbol{\Lambda}_{{\mathrm{\bf{H}}}} \in \mathbb{C}^{4 \times 4}$ is a diagonal matrix.}
If ${{\bf{H}}}$ is known by the transmitter, then, from \cite[Prop. 2]{Xiao2011TSP}, the optimal design satisfies
$\mathbf{U}_{{\mathrm{\bf{B}}}} = \mathbf{V}_{{\mathrm{\bf{H}}}}^H$ and (\ref{eq:y_4_exp}) becomes
\begin{equation}\label{eq:y_4_exp_2}
\begin{array}{l}
\overline {\bf{y}} =
\left[ {\begin{array}{*{20}{c}}
{{a_1\lambda_1}}&{}&{}\\
{}& \ddots &{}\\
{}&{}&{{a_4 \lambda_4}}
\end{array}} \right] \! \! \left[ {\begin{array}{*{20}{c}}
{{V_{11}}}& \hdots &{{V_{14}}}\\
\vdots & \ddots & \vdots \\
{{V_{41}}}&\hdots&{{V_{44}}}
\end{array}} \right]   \mathbf{d} + {{\bf{n}}}
\end{array}
\end{equation}
where $\overline {\bf{y}} = \mathbf{U}_{{\mathrm{\bf{H}}}}^H {\bf{y}}$ while $a_i$
and $\lambda_i$ are the diagonal entries of $\boldsymbol{\Lambda}_{{\mathrm{\bf{H}}}}$
and $\boldsymbol{\Lambda}_{\mathrm{\bf{B}}}$, respectively, and $V_{ij} =\left[\mathbf{V}_{\mathrm{\bf{B}}} \right]_{ij}$.

Suppose that two of the subchannel gains, say $a_2$ and $a_4$, are very weak.
Then, with a Gaussian-signal precoder, the powers allocated to the corresponding subchannels will be
very small even at moderate SNRs. Since, with Gaussian signals, ${{\bf{V}}_{\mathrm{\bf{B}}}}$ is {\bl an identity matrix}, $d_2$ and $d_4$ are essentially muted.
With a proper ${{\bf{V}}_{\mathrm{\bf{B}}}}$, in contrast, the received signal satisfies
\begin{equation}\label{eq:y_4_exp_3}
{\left[ {\overline {\bf{y}} } \right]_i} = {a_i}{\lambda _i}\sum\limits_{j = 1}^4 {{V_{ij}}{d_j}} \qquad i = 1,2,3,4
\end{equation}
and now, even if $a_2 \lambda_2 \approx 0$ and $a_4 \lambda_4 \approx 0$, $d_2$ and $d_4$ can still
be effectively transmitted along other subchannels.
%If we also allocate the power $\lambda_i$ based on the mutual information with finite alphabet inputs, then
%we can completely compensate the performance loss caused by the Gaussian input design.
%Iterative algorithms can be designed to find the optimal value of $\lambda_i$  and $V_{ij}$.
\end{example}

\vspace{2mm}

As indicated by (\ref{eq:y_4_exp_3}), an adequate design for discrete constellations generally mixes
all the signals ($d_1, d_2, d_3, d_4$) and transmits the ensuing beams on different subchannels.
{\bl This is referred as a complete search design.} {\bl In fact, for single-user MIMO systems with finite-alphabet inputs,
a complete search design can achieve the maximal mutual information and near-maximal mutual information with instantaneous CSI \cite{Xiao2011TSP}
and statistical CSI \cite{zeng2012linear} at the transmitter, respectively.
However, the search space grows exponentially with $N_{\mathrm t}$ \cite{Xiao2011TSP}.}
% for computing the mutual information with finite alphabet inputs

%Therefore, when $N_{\mathrm t}$ becomes large, the computational complexity of the conventional design is prohibitive.

Intuitively though, if there are two weak subchannels, say $a_2$ and $a_4$ in Example 1, it is not necessary to mix all the signals.
It suffices to mix $d_2$ with $d_1$, and $d_4$ with $d_3$, and then transmit the ensuing beams on the strong subchannels $a_1$ and $a_3$.
This corresponds to
\begin{equation}\label{eq:y_4_exp_4}
{\bf{V}}_{\mathrm{\bf{B}}} = \left[ {\begin{array}{*{20}{c}}
{{V_{11}}}&{{V_{12}}}&0&0\\
{{V_{21}}}&{{V_{22}}}&0&0\\
0&0&{{V_{33}}}&{{V_{34}}}\\
0&0&{{V_{43}}}&{{V_{44}}}
\end{array}} \right]
\end{equation}
which, plugged into (\ref{eq:y_4_exp_2}), gives
\begin{align}
{\left[ {\overline {\bf{y}} } \right]_i} & = {a_i}{\lambda _i}\sum\limits_{j = 1}^2 {{V_{ij}}{d_j}} \qquad  i = 1,2 \label{eq:y_4_exp_5} \\
{\left[ {\overline {\bf{y}} } \right]_i} & = {a_i}{\lambda _i}\sum\limits_{j = 3}^4 {{V_{ij}}{d_j}} \qquad  i = 3,4 .  \label{eq:y_4_exp_6}
\end{align}
Observe from (\ref{eq:y_4_exp_5}) and (\ref{eq:y_4_exp_6}) that
$(d_1,d_2)$ and $(d_3,d_4)$ are decoupled. {\bl This is referred as a per-group search design.}
If the entries of $\mathbf{d}$ are QPSK, then the search space
%for computing the mutual information with finite alphabet inputs in (\ref{eq:y_4_exp_5}) and (\ref{eq:y_4_exp_6})
is of dimension $2 \times 4^{2 \times 2} = 512$ \cite{Xiao2011TSP}.  In contrast, for the complete search in (\ref{eq:y_4_exp_3}), it would be of dimension $ 4^{2 \times 4} = 65536$. As will be seen, this enormous reduction in complexity may incur only a minute loss in performance.
%We know from (\ref{eq:y_4_exp_5}) and (\ref{eq:y_4_exp_6}) that by paring a strong subchannel with a weak subchannel, i.e., $a_1$ and $a_2$, $a_3$ and $a_4$,  $d_2$  and $d_4$ are transmitted all the same. Therefore, the structure in (\ref{eq:y_4_exp_4}) may perform close to the complete-search design, but with a substantially lower computational complexity.

With instantaneous CSI, the idea suggested in Example \ref{4_4} leads to the PGP technique in \cite{Ketseoglou2015TWC}.
%However, the discussion in Example \ref{4_4} provides a clear instruction that pairing strong subchannels with weak subchannels can reduce the computational complexity significantly while maintaining  a close performance to the complete-search design, which is  missing in \cite{Ketseoglou2015TWC}. This instruction is important for designing the low-complexity algorithm for the precoder optimization.
A more general construction that does not require instantaneous CSI at the transmitter is presented next.
{\bl Table \ref{table:precoder} provides a comparison between the previous work for precoder designs
for single-user MIMO with finite-alphabet inputs and the work in this paper. }

\begin{table}
{\bl
\centering
  \renewcommand{\multirowsetup}{\centering}
 \captionstyle{center}
  {
\caption{Precoder designs for single-user MIMO channel with finite alphabet inputs}
\vspace{0.2cm}
\label{table:precoder}
\begin{tabular}{|c|c|c|c|}
\hline
  Paper   &  CSI at Transmitter  &  Precoder Structure &   Performance \\ \hline
C. Xiao \textit{et al.} \cite{Xiao2011TSP} & Instantaneous CSI & Complete search & Optimal  \\ \hline
\multirow{2}{*} {W. Zeng \textit{et al.} \cite{zeng2012linear}} & Statistical CSI  & \multirow{2}{*} {Complete search}  & \multirow{2}{*} {Near-optimal}  \\
 & Kronecker fading &   & \\ \hline
S. K. Mohammed \textit{et al.} \cite{Mohammed2011TIT} & Instantaneous CSI & Per-group search with fixed $\boldsymbol{\Lambda}_{\mathrm{\bf{B}}}$ and ${{\bf{V}}_{\mathrm{\bf{B}}}}$ & Suboptimal  \\ \hline
T. Ketseoglou \textit{et al.} \cite{Ketseoglou2015TWC} & Instantaneous CSI & Per-group search with optimized $\boldsymbol{\Lambda}_{\mathrm{\bf{B}}}$ and ${{\bf{V}}_{\mathrm{\bf{B}}}}$ & Near-optimal  \\ \hline
 \multirow{2}{*} {The work in this paper} & Statistical CSI  &  \multirow{2}{*} {Per-group search with optimized $\boldsymbol{\Lambda}_{\mathrm{\bf{B}}}$ and ${{\bf{V}}_{\mathrm{\bf{B}}}}$} &  \multirow{2}{*} {Near-optimal}  \\
 &  Jointly correlated Rician fading &  &  \\  \hline
\end{tabular}
}
}
\end{table}

\section{Low-Complexity Precoder Design}

%In this section, we investigate low-complexity MIMO precoding for finite alphabet signals.
%First, we introduce the channel model. Then, we provide an asymptotic (large system limit)
%expression of the erogdic mutual information in (\ref{eq:Mutual_Info}). Based on this asymptotic expression, we formulate
%precoder structures, from which a low-complexity numerical algorithm is proposed to design the precoder.

\subsection{Channel Model}
\label{sta:model}

%Inspired by (\ref{eq:y_4_exp_5}) and (\ref{eq:y_4_exp_6}), we propose a low-complexity
%design to maximize the spectral efficiency.
To avoid modeling artifacts in the design of the precoder, we consider the rather general MIMO channel model
%In the jointly-correlated model, the correlation between the transmitter and receiver side is modeled jointly. Specifically, the channel matrix $\qH$ is given by
\begin{equation}\label{eq:Spatial_Cov}
\qH = \qU_{{\rm R}} \left( \tilde{\qG} \odot \qW \right) \qU_{{\rm T}}^{H} + \bar\qH
\end{equation}
where $\qU_{{\rm R}} = \left[\qu_{{\rm R},1}, \qu_{{\rm R},2},\ldots, \qu_{{\rm R}, N_{\mathrm r}} \right] \in\mathbb{C}^{N_{\mathrm r}\times N_{\mathrm r}}$ and $\qU_{{\rm T}} =\left[\qu_{{\rm T},1}, \qu_{{\rm T},2},\ldots,\qu_{{\rm T},N_{\mathrm t}} \right]
\in\mathbb{C}^{N_{\mathrm t} \times N_{\mathrm t}}$ are deterministic unitary matrices, $\tilde{\qG}$ is a deterministic matrix of size $N_{\mathrm r} \times N_{\mathrm t}$ with real-valued nonnegative entries, $\qW
\in\mathbb{C}^{N_{\mathrm r} \times N_{\mathrm t}}$ is a random matrix whose entries are independent and identically distributed (IID) complex Gaussian with zero-mean and unit-variance, and $\bar\qH\in\mathbb{C}^{N_{\mathrm r} \times N_{\mathrm t}}$
is a deterministic matrix modeling the Rice component. We further define
${\bf{G}}  = {\bf{\tilde{{G}}}}  \odot {\bf{ \tilde{{G}}}}$ such that $\left[\qG\right]_{nm}$ is the average power coupling
between ${\bf{u}}_{{\rm{R}},n}$ and ${\bf{u}}_{{\rm{T}},m}$ \cite{weichselberger2006stochastic}.  The transmit and
receive correlation matrices of $\qH$ are
\begin{equation}\label{correlation}
\begin{array}{l}
 {\bf{R}}_{\mathrm t}  = E_{{\bf{H}}}\left[ { \left({\bf{H} - {\bar\qH}}\right)^H \left({\bf{H} - \bar\qH} \right)} \right] =  {\bf{U}}_{{\rm{T}}} \boldsymbol{\Gamma} _{{\rm{T}} } {\bf{U}}_{{\rm{T}} }^H  \\
 {\bf{R}}_{\mathrm r}  = E_{{\bf{H}}}\left[ {  \left({\bf{H}} - {\bar\qH} \right) \left({\bf{H}} - {\bar\qH} \right)^H } \right] = {\bf{U}}_{{\rm{R}}} \boldsymbol{\Gamma} _{{\rm{R}} } {\bf{U}}_{{\rm{R}} }^H  \\
 \end{array}
\end{equation}
where  $\boldsymbol{\Gamma}_{{\rm{T}}} $ and  $\boldsymbol{\Gamma}_{{\rm{R}}}$ are diagonal with
$\left[ {{\bf{\Gamma }}_{{\rm{T}}} } \right]_{mm}  = \sum\nolimits_{n = 1}^{N_{\mathrm r} } { \left[\qG\right]_{nm} }$, for $m = 1,2,\ldots,N_{\mathrm t}$, and $
\left[ {{\bf{\Gamma }}_{{\rm{R}}} } \right]_{nn}  = \sum\nolimits_{m = 1}^{N_{\mathrm t} } { \left[\qG\right]_{nm} }$, for $n = 1,2,\ldots,N_{\mathrm r}$, respectively.

We note that (\ref{eq:Spatial_Cov}) subsumes most statistical MIMO
channel models. For instance, if $\bar\qH = \mathbf{0}$ and $\mathbf{G}$ is rank-one, the Kronecker model is recovered\cite{shiu2000fading,Kermoal2002JSAC,xiao2004TWC}. Allowing $\mathbf{G}$ to have arbitrary rank
while fixing ${\bf{U}}_{{\rm{R}}}$ and ${\bf{U}}_{{\rm{T}}}$ to be
Fourier matrices, we obtain the virtual channel
representation for uniform linear arrays (ULA)\cite{sayeed2002deconstructing}.
If we further relax ${\bf{U}}_{{\rm{R}}}$ and ${\bf{U}}_{{\rm{T}}}$
to be arbitrary unitary matrices, we obtain the Weichselberger's
channel model \cite{weichselberger2006stochastic}.
As far as the Rice component is concerned, and in contrast with works where its structure is restricted \cite{Veeravalli2005TIT,Gao}, in our model it is also arbitrary.

%For the characterization of the LOS effect, the virtual
%channel representation for the ULA MIMO channels in \cite{Veeravalli2005TIT} includes one LOS component.
%The jointed-correlated MIMO channel model in  \cite{Gao} includes multiple LOS components.  In \cite{Gao},
%$\bar\qH $ is modeled as a deterministic
%matrix with at most one nonzero element in each row and each column.
%Here, we allow $\bar\qH$ to be arbitrary structure to cover more
%general scenarios, such as instantaneous CSI and imperfect CSI
%cases.

Without loss of generality, we normalize $\qG$ and $\bar\qH$ such that
\begin{align}
\label{eq:PowNorm}
\frac{1}{N_{\mathrm r} N_{\mathrm t}} \|\qG\|_F &=\frac{1}{K} \\
\frac{1}{N_{\mathrm r} N_{\mathrm t}} \| \bar\qH \|^2_F &=\frac{K}{K+1}
\end{align}
where $K$ is the Rice factor.  For $K \rightarrow\infty$ and $K =0$, (\ref{eq:Spatial_Cov})
reduces to a deterministic channel and  a Rayleigh-faded channel, respectively.

In this work, we assume  that the receiver knows $\qH$ perfectly whereas the transmitter only
has statistical knowledge thereof, i.e., the transmitter only knows $\bar\qH$, $\qU_{{\rm R}}$, $\tilde{\qG}$,
and $\qU_{{\rm T}}$. {
As indicated in \cite[Table II]{Ahumada2005TVT}, the coherence time of the channel statistics exceeds 1 s in
typical residential urban environments\footnote{\bl  Measurements for a single-input single-output (SISO) narrowband system were presented in \cite{Ahumada2005TVT}. In general, the channel coherence time  is mainly determined by the velocity of the user and the carrier frequency \cite[Eq. (5.40)]{Rappaport2002}. The number of transmit and receive antennas has little impact on the channel coherence time. Moreover,
it is proved in \cite[Prop. 1]{Liu2003TIT} that the channel statistics is independent of the frequency for a wideband system.
Therefore, for the coherence time of the channel statistics, there is no  obvious difference
between a SISO narrowband system and a MIMO wideband system.}.
 The Long Term Evolution (LTE) specification defines a subframe as a transmission time interval of 1 ms \cite{Furht}. Therefore, once
 $\bar\qH$, $\qU_{{\rm R}}$, $\tilde{\qG}$, and $\qU_{{\rm T}}$ are obtained and fed back to the transmitter, they can be used
 for hundreds of subframes. As a result, the overall feedback overhead for precoder designs that are based on
 statistical CSI is much smaller than that of precoder designs requiring instantaneous CSI\footnote{ For precoder designs requiring instantaneous CSI, the feedback overhead can also be reduced by exploiting  vector quantization\cite{Rajashekar2016TCom}.}. }

% in (\ref{eq:Spatial_Cov}).

With $\qH$ known at the receiver, the ergodic mutual information between $\qx$ and $\qy$  is given by \cite{Cover}
\begin{equation}\label{eq:Mutual_Info}
I(\qx;\qy)=E_{\bl \mathbf{H}} \left[E_{\bl \mathbf{x},\mathbf{y}} \left[\left.\log\frac{p (\qy|\qx,\qH)} {p(\qy|\qH)}\right|\qH\right]\right]
\end{equation}
where the outer expectation is over $\qH$ and the inner expectation is over $p(\qx,\qy|\qH)$.

\subsection{Mutual Information in the Large-Dimensional Regime}
{
The ergodic mutual information in (\ref{eq:Mutual_Info})
requires the expectation with respect to the distribution of $\qH$,
which can not be obtained in  closed form.
To overcome this problem, the concept of the deterministic equivalent channel \cite{Wu2015TWCOM}
can be exploited to approximate
(\ref{eq:Mutual_Info}) in the large-dimensional regime.
Using the deterministic equivalent channel, we can
then obtain the counterparts to (\ref{eq:y_4_exp_5}) and (\ref{eq:y_4_exp_6}) for the general setting.
To this end, we assume that both $N_{\mathrm r}$ and $N_{\mathrm t}$ grow
large with ratio $c = N_{\mathrm t}/N_{\mathrm r}$.
In the following, we define this deterministic equivalent channel and the parameters
used to compute its mutual information.
}

Let us define the vector channel
\begin{equation}\label{eq:EqScalGAUEach0}
\qz= \qXi^{1/2} \qx+ \check{\bf n}
\end{equation}
where {$\qXi$ is the deterministic equivalent channel matrix used to approximate the exact
ergodic mutual information in (\ref{eq:Mutual_Info})} and $\check{\bf n} \in \bbC^{N_{\mathrm t} \times 1} $ is a standard complex Gaussian random vector. The  minimum mean-square error (MMSE) estimate of $\qx$ based on the observation of $\qz$ is
\begin{equation}\label{eq:hatx_k}
 \hat\qx (\qz) = E \big [\qx |\qz \big]
\end{equation}
where the expectation is over $p(\qx|\qz)$.
The covariance of the estimation error is the MMSE matrix \cite{Xiao2011TSP,Perez-Cruz2010TIT,Palomar2006TIT}
\begin{equation} \label{eq:Omega}
    \qOmega = E \left[ \big(\qx - \hat\qx(\qz)\big) \big(\qx - \hat\qx(\qz)\big)^H \right]
\end{equation}
with expectation over $\qx$ and $\qz$.

Next, we introduce several useful quantities. Define $\boldsymbol{\gamma} = [\gamma_{1},\gamma_{2}, \dots, \gamma_{N_{\mathrm r}}]^T$, $\boldsymbol{\psi} = [\psi_{1}, \psi_{2}, \dots, \psi_{N_{\mathrm t}}]^T$, and
\begin{align}
    \qXi  = \qT + \bar{\qH}^H \left(\qI_{N_{\mathrm r}}+ \qR\right)^{-1} \bar{\qH} \qquad \in\mathbb{C}^{N_{\mathrm t} \times N_{\mathrm t}}. \label{eq:eqChxi}
\end{align}
The equivalent channel  matrix $\qXi$ is a function of the
auxiliary variables $\{ \boldsymbol{\gamma}, \boldsymbol{\psi}, {\mathbf R}, {\mathbf T}\}$,
which satisfy the coupled equations
\begin{align} \label{eq:eqChMatrixTR}
\qT &  = \qU_{{\rm T}}^H {\rm{diag}}\left( \qG^T \boldsymbol{\gamma} \right)\qU_{{\rm T}} \qquad \in\mathbb{C}^{N_{\mathrm t} \times N_{\mathrm t}}\\
\qR & = \qU_{{\rm R}}^H {\rm{diag}}\left( \qG \boldsymbol{\psi}\right)\qU_{{\rm R}} \qquad\;\; \in\mathbb{C}^{N_{\mathrm r} \times N_{\mathrm r}}
\end{align}
while the entries of $\boldsymbol{\gamma}$ and  $\boldsymbol{\psi}$
are the solutions to the fixed-point equations
\begin{align}
\gamma_{m} &=  \qu_{{\rm R},m}^{H} \left(\qI_{N_{\mathrm r}}+\qR\right)^{-1} \qu_{{\rm R},m}  - \qu_{{\rm R},m}^{H} \left(\qI_{N_{\mathrm r}} + \qR\right)^{-1} \bar{\qH} \, \qOmega \, \bar{\qH}^{H} \left(\qI_{N_{\mathrm r}}+ \qR\right)^{-1}\qu_{{\rm R},m}  \label{eq:Varsigma_k-MSE} \\
\psi_{n}   &= \qu_{{\rm T},n}^{H} \, \qOmega \, \qu_{{\rm T},n}  \label{eq:Varsigma_k-MSE_2}.
\end{align}
{\bl The equivalent channel matrix $\qXi $ in (\ref{eq:EqScalGAUEach0})
does not depend on the instantaneous channel realizations, but it is merely an instrument to obtain an
asymptotic expression for the ergodic mutual information in (\ref{eq:Mutual_Info}), which is given as follows. }

\begin{prop}\label{Proposition_1}
In the large-dimensional regime, the mutual information in (\ref{eq:Mutual_Info}) satisfies
\begin{equation}\label{eq:GAUMutuall}
I(\qx;\qy) \simeq I_{\rm asy}(\qx;\qy)
\end{equation}
where
\begin{equation}\label{eq:GAUMutuall_2}
 I_{\rm asy}(\qx;\qy) = I\left( \qx;\qz \right) + \log\det\left(\qI_{N_{\mathrm r}} + \qR\right)-  \boldsymbol{\gamma}^T \qG \boldsymbol{\psi}
\end{equation}
with $ I (\qx;\qz )$ being the mutual information over the equivalent channel in (\ref{eq:EqScalGAUEach0}).
The approximation in (\ref{eq:GAUMutuall}) sharpens as the matrices become large.

\begin{proof}
See Appendix \ref{Proof_Proposition_1}.
\end{proof}
\end{prop}

\vspace{2mm}

{\bl
{\textit{Remark 1:}} We note that there are three main differences between the asymptotic expression in Proposition 1 and the asymptotic expression in \cite{Hachem2006AAP}. First, our asymptotic expression and the asymptotic expression in \cite{Hachem2006AAP}
apply for the mutual information with finite alphabet inputs and  Gaussian inputs, respectively.
Therefore, the employed mathematical methods are completely different.  The derivation of the asymptotic expressions
relies  the replica method and the Stieltjes transform for finite alphabet inputs and  Gaussian
inputs, respectively. Second, our expression applies for correlated fading channels while the expression
in \cite{Hachem2006AAP} only applies for  independent fading channels. Third, our expression accounts for the Rician
factor.
}

%{\textit{Remark 1:}}
%It should be noted that the asymptotic expression in (\ref{eq:GAUMutuall}) is general and applies to arbitrary statistical CSI matrices  $\bar\qH$,  $\qU_{{\rm R}}$, $\tilde{\qG}$, $\qU_{{\rm T}}$, and arbitrary transmit signal distributions $\qx$.
%When the dimensions of the matrices are finite, (\ref{eq:GAUMutuall}) approximates $I(\qx;\qy)$ with an accuracy that improves as the dimensions become large. In addition,
%For $K\rightarrow\infty$, $\qH$ becomes deterministic and (\ref{eq:GAUMutuall}) naturally reduces to the exact mutual information under instantaneous CSI.
 In the following, we shall take advantage of the asymptotic mutual information expression in Proposition 1 to design the precoder $\qB$.

\subsection{Precoder Structure}
\subsubsection{Structure of $\mathbf{U}_{{\mathrm{\qB}}}$}

Consider the eigenvalue decomposition $\qXi = \mathbf{U}_{{{\qXi}}} {\bf{\Lambda}}_{\qXi} \mathbf{U}_{\qXi}^H$ where ${\bf{\Lambda}}_{\qXi} \in \mathbb{C}^{N_{\mathrm{t}} \times N_{\mathrm{t}}}$ is diagonal and
$\mathbf{U}_{{{\qXi}}} \in \mathbb{C}^{N_{\mathrm{t}} \times N_{\mathrm{t}}}$ is unitary.
%Then, we have the following proposition.

\vspace{2mm}

\begin{prop}\label{prop:opt_structure}

The precoder left singular matrix $\mathbf{U}_{{\mathrm \qB}}$ that maximizes the asymptotic mutual
information in (\ref{eq:GAUMutuall}) equals $\mathbf{U}_{\qXi}$.

\begin{proof}
See Appendix \ref{Proof_prop:opt_structure}.
\end{proof}
\end{prop}

\vspace{2mm}

This result generalizes what was found in \cite{zeng2012linear} for Kronecker channels, where it is optimal to transmit along the eigendirections of the transmit correlation matrix $\mathbf{U}_{{\rm T}}$.

%{\textit{Remark 1:}} For the precoder design with statistical CSI and finite
%alphabet input constraints, the optimal precoder structure of the left singular matrix
%has been proved to be $\mathbf{U}_{{\rm T}}$ for the special case of the Kronecker fading model \cite{zeng2012linear}.
%However, Proposition \ref{prop:opt_structure} indicates that if we allow  $\bar{\qH}$ to be arbitrary structure, aligning the signal
%along the eigen-direction of the transmit correlation matrix $\mathbf{U}_{{\rm T}}$ is no longer
%optimal.

Now, plugging $\mathbf{U}_{{\rm \qB}} = \mathbf{U}_{\qXi}$ into (\ref{eq:EqScalGAUEach0}) and using \cite[(8)]{Xiao2011TSP}, we can rewrite (\ref{eq:EqScalGAUEach0}) as

\begin{equation}\label{eq:EqScalGAUEach}
\qz_{\rm eq} =  {\bf{\Lambda}}_{\qXi}^{1/2} \qx_{\rm eq} + \check{\bf n}
\end{equation}
where
\begin{align}
{{\bf{z}}_{{\rm{eq}}}} & =  \qU_{\qXi}^H \qz  \label{eq:z_eq}\\
\qx_{\rm eq} & =  { {{\bf{\Lambda }}_{{{\mathrm{\bf{B}}}}}}{{\bf{V}}_{{{\mathrm{\bf{B}}}}}} } \, \qd .   \label{eq:x_eq}
\end{align}

Let us divide the transmit signal $\qd$ into $S$ streams.
Each stream $\qd_{\mathrm s} \in \mathbb{C}^{N_{\mathrm s} \times 1}$ is to be conveyed over $N_{\mathrm s}=N_{\mathrm t}/S$ diagonal entries of ${\bf{\Lambda}}_{\qXi} $.
Let the set $\left\{\ell_{1},\ldots,\ell_{N_{\mathrm t}}\right\}$ denote
a permutation of $\left\{1,\ldots,N_{\mathrm t}\right\}$ and
let ${\bf{\Lambda }}_s \in \mathbb{C}^{N_{\mathrm s} \times N_{\mathrm s}}$
and $\qV_s \in \mathbb{C}^{N_{\mathrm s} \times N_{\mathrm s}}$ denote a diagonal matrix
and a unitary matrix, respectively, for $s = 1,\ldots,S$.
${\bf{\Lambda }}_s $ and $\qV_s$ will be optimized later.
The goal of arranging these $S$ streams as in (\ref{eq:y_4_exp_5})
and (\ref{eq:y_4_exp_6}) prompts the subsequent design steps.

\subsubsection{Structure of ${\bf{\Lambda}}_{\mathrm{\bf{B}}}$}

We set
\begin{align}\label{eq:Lambda_pair}
{\left[ {{{\bf{\Lambda }}_{{{\mathrm{\bf{B}}}}}}} \right]_{\ell_{j}\ell_{j}}} = \left[ {{{\bf{\Lambda}}_{s}}} \right]_{ii}
\end{align}
where $i = 1,\ldots,N_{\mathrm s}$, $s=1,\ldots,S$ and $j = (s - 1) N_{\mathrm s} + i$.
With this structure, the $s$th stream is transmitted along the $\ell_{(s - 1) N_{\mathrm s} + 1},\ldots, \ell_{(s - 1) N_{\mathrm s} + N_{\mathrm s}}$ diagonal entries of ${\bf{\Lambda}}_{\qXi}$.

\subsubsection{Structure of ${{\bf{V}}_{{{\mathrm{\bf{B}}}}}}$}

Here we set
\begin{align}\label{eq:V_pair}
 \left[ {{{\bf{V}}_{{{\mathrm{\bf{B}}}}}}} \right]_{\ell_i \ell_j} =   \left\{ \begin{array}{l}
   {\left[ {{{\bf{V}}_{s}}} \right]_{mn}} \qquad i = (s - 1) N_{\mathrm s} + m , \ j = (s - 1) N_{\mathrm s}  + n   \\
0 \qquad \qquad\;\; {\rm otherwise}
 \end{array} \right.
\end{align}
for $m = 1,\ldots, N_{\mathrm s}$, $n = 1,\ldots,N_{\mathrm s}$, $s = 1,\ldots, S$,
$i =  1,\ldots, N_{\mathrm t}$, and $j = 1,\ldots, N_{\mathrm t}$. With this structure,
the $s$th stream is mapped only to rows $\ell_{(s - 1) N_{\mathrm s} + 1},\ldots, \ell_{(s - 1) N_{\mathrm s}+ N_{\mathrm s}}$ and columns $\ell_{(s - 1) N_{\mathrm s}+ 1},\ldots, \ell_{(s - 1) N_{\mathrm s} + N_{\mathrm s}}$ of  ${{{\bf{V}}_{{{\mathrm{\bf{B}}}}}}} $.
This yields $S$ decoupled groups of streams at the receiver.
%This decouples $S$ streams at the receiver. NOT CLEAR. DO YOU MEAN THAT S GROUPS OF STREAMS ARE DECOUPLED, OR THAT GROUPS OF S STREAMS ARE DECOUPLED?

The design in (\ref{eq:y_4_exp_4}) is a specific instance of
(\ref{eq:V_pair}) with $\left\{\ell_{1},\ldots,\ell_{N_{\mathrm t}}\right\}
 = \left\{1,2,3,4\right\}$ and $S =2$. Recall how $(d_1,d_2)$ and
 $(d_3,d_4)$ are indeed decoupled in (\ref{eq:y_4_exp_5}) and (\ref{eq:y_4_exp_6}).

\subsubsection{Structure of $\qd_{s}$}

Finally, we let
\begin{align}\label{eq:d_pair}
\left[\qd_{s}\right]_{i} =  \left[\qd \right]_{\ell_{j}}
\end{align}
where $i = 1,2,\ldots,N_{\mathrm s}$, $s = 1,2,\ldots, S$, and $j = (s - 1)N_{\mathrm s} + i$.

\subsection{Precoder Optimization}

Based on (\ref{eq:Lambda_pair})--(\ref{eq:d_pair}), the relationship in (\ref{eq:x_eq}) becomes
\begin{align}\label{eq:x_eq_pair}
 \left[\qx_{\rm eq}\right]_{\ell_{j}} =  \left[{\bf{\Lambda }}_{{s}}  {\bf{V}}_{{s}}  \qd_{{s}}\right]_{i}
\end{align}
for $i = 1,\ldots,N_{\mathrm s}$, $s = 1,\ldots, S$, and  $j = (s - 1) N_{\mathrm s} + i$.
Recalling that
${\bf{\Lambda}}_{\mathrm \qXi}$ is diagonal, (\ref{eq:EqScalGAUEach}) then reduces to
\begin{align}\label{eq:z_eq_pair}
 \left[\qz_{\rm eq}\right]_{\ell_{j}} & = \left[ {\bf{\Lambda}}_{\qXi} \right]_{\ell_{j}\ell_{j}}^{1/2}  \left[\qx_{\rm eq}\right]_{\ell_{j}} + [{\mathbf n}_s]_{i}
\end{align}
where $ [{\mathbf n}_s]_{i} =  [\check{\mathbf n}]_{\ell_{j}}$.
%where $i = 1,2,\ldots,N_{\mathrm s}$,  $s = 1,2,\ldots,S$, and $j = (N_{\mathrm s} - 1)s + i$.

Equations (\ref{eq:x_eq_pair}) and (\ref{eq:z_eq_pair}) indicate that each independent data stream
$\qd_{s} $ is transmitted along its own $N_{\mathrm s}$ separate subchannels without
interfering with other streams.
Furthermore,  the MMSE matrix then equals
\begin{align}\label{eq:Omega_pair}
 {\left[  \qOmega\right]_{\ell_i \ell_j}} =   \left\{ \begin{array}{lll}
{\left[ \qOmega_{s}  \right]_{mn}} & \qquad i = (s - 1) N_{\mathrm s} + m , \ j = (s - 1) N_{\mathrm s} + n   \\
0 & \qquad {\rm otherwise}   \\
\end{array} \right.
\end{align}
%where $ m = 1,2,\ldots, N_{\mathrm s} $, $ n = 1,2,\ldots,N_{\mathrm s}$, $s = 1,2,\ldots, S$, $i =  1,2,\ldots, N_{\mathrm t}$, and $j =  1,2,\ldots, N_{\mathrm t}$.
where
\begin{equation}\label{eq:MMSE_eq}
\qOmega_{s}  = {{\bf{\Lambda }}_{s}}{{\bf{V}}_{s}} \, E\left[ {\big( { {{\bf{d}}_{s}} - {{{\bf{\hat d}}}_{s}}} \big){{\big( {{{\bf{d}}_{s}} - {{{\bf{\hat d}}}_{s}} } \big)}^H}} \right] {\bf{V}}_{s}^H{\bf{\Lambda }}_{s}^H
\end{equation}
with
\begin{equation}\label{eq:q}
{{{\bf{\hat d}}}_{s}} = E\big[ {{\bf{d}}_{s}} |\qz_{s} \big]
\end{equation}
and $\left[\qz_{s}\right]_{i} =  \left[\qz_{\rm eq}\right]_{\ell_{j}}$, and further defining diagonal matrices $\qXi_s$, for $s=1,\ldots,S$, with entries $\left[\qXi_{s}\right]_{ii} = \left[
{\bf{\Lambda}}_{\qXi}  \right]_{\ell_{j}\ell_{j}}$.
%$i = 1,2,\ldots,N_{\mathrm s}, s = 1,2,\ldots,S, j = (N_{\mathrm s} - 1)s + i$.

The main term within (\ref{eq:GAUMutuall_2}) can now be expressed as
\begin{equation}\label{eq:I_pair}
I\left( \qx ;\qz \right) = \sum\limits_{s = 1}^S I\left( {{\bf{d}}_{s}};\qz_{s}  \right)
\end{equation}
based on which the gradients of $ I_{\rm asy}(\qx;\qy)$ with respect to $\mathbf{\Lambda}_{s}^2$ and $\mathbf{V}_{s}$
are given by \cite[(22)]{Palomar2006TIT},
\begin{align}
\label{eq:gamma_gradient}
{\nabla _{{\bf{\Lambda }}_{s}^2}} I_{\rm asy}(\qx;\qy) & =  \diag\left( {{\bf{V}}_{s}^H{{\bf{E}}_{s}}{{\bf{V}}_{s}}} \qXi_{s} \right) \\
\label{eq:V_gradient}
{\nabla _{{{\bf{V}}_{s}}}} I_{\rm asy}(\qx;\qy) &  =   \qXi_{s} {\bf{\Lambda }}_{s}^2{{\bf{V}}_{s}}{{\bf{E}}_{s}}
\end{align}
where
\begin{equation}\label{eq:Es}
{{\bf{E}}_{s}} = E\left[ {\left( { {{\bf{d}}_{s}} - {{{\bf{\hat d}}}_{s}}} \right){{\left( {{{\bf{d}}_{s}}
- {{{\bf{\hat d}}}_{s}} } \right)}^H}} \right].
\end{equation}

\vspace{2mm}

Based on Propositions \ref{Proposition_1} and \ref{prop:opt_structure}, on (\ref{eq:I_pair}), and on the relationship between $\mathbf{\Lambda}_{1},\ldots,\mathbf{\Lambda}_{S}$ and ${{{\bf{\Lambda }}_{{{\mathrm{\bf{B}}}}}}}$ in (\ref{eq:Lambda_pair}) as well as the relationship between
$\mathbf{V}_{1},\ldots,\mathbf{V}_{S}$ and ${{{\bf{V}}_{{{\mathrm{\bf{B}}}}}}}$ in (\ref{eq:V_pair}),
we propose Algorithm \ref{Gradient_Pair} to optimize $\qB$.
In Steps 3 and 5 of this algorithm, ${\mathbf{\Lambda}}_{{s}}^{(n)}$ and $\mathbf{V}_{{s}}^{(n)}$
are updated along the gradient descent direction, with the backtracking line search method used to determine the step size.
In Step 4, ${\mathbf{\Lambda}}_{{s}}^{(n)}$ is normalized to satisfy the power constraint.
In Step 6, $\qXi$, ${\bf{R}}$,
 $\gamma$, and $\psi$ are updated for the new precoder based on
(\ref{eq:eqChxi})--(\ref{eq:Varsigma_k-MSE_2}), (\ref{eq:Omega_pair}).
In Step 7, if  $n$ is less than some maximum number of iterations
and $I^{(n + 1)} ( \qx;\qy )  - I^{(n )} ( \qx;\qy )$
is above some threshold, the iterations continue; otherwise, the algorithm is stopped. In Step 8, we compute the optimal ${{{\bf{U}}_{{{\mathrm{\bf{B}}}}}}}$
based on Proposition \ref{prop:opt_structure}.

With statistical CSI, the expectation of the mutual information in (\ref{eq:Mutual_Info})
can be evaluated efficiently by applying \cite[Prop. 2]{zeng2012linear}.
Likewise, operations such as  matrix products and the fixed-point equations are polynomial functions of the numbers of antennas, and thus can also be performed efficiently. What dominates the computational cost is expecting the mutual information and the MMSE matrix over $\mathbf{d}_{m}$, as the complexity
of these expectations is exponential in $N_{\mathrm t}$ \cite[(14) and (47)]{zeng2012linear}. \,
Therefore, it suffices to

\begin{alg} \label{Gradient_Pair}
Maximization of $I\left( \qx;\qy \right) $ with respect to $\qB$.

\vspace*{1.5mm} \hrule \vspace*{1mm}
  \begin{enumerate}

\itemsep=0pt

\item Initialize $\mathbf{\Lambda}_{{s}}^{(0)}$, $\mathbf{V}_{{s}}^{(0)}$ for  $s = 1,\ldots,S$.
%as well as $\qXi_{\rm eq}$, ${\bf{R}}_{\rm eq}$, $\gamma_{\rm eq}$ and $\psi_{\rm eq}$.
Fix a maximum number of iterations, $N_{\rm iter}$, and a threshold $\varepsilon$.

\item  Initialize $\qXi$, ${\bf{R}}$,
 $\boldsymbol{\gamma}$, and $\boldsymbol{\psi}$
based on (\ref{eq:eqChxi})--(\ref{eq:Varsigma_k-MSE_2}), with $\qOmega$
based on (\ref{eq:Omega_pair}).  Then, initialize $I^{\left(1\right)}\left( \qx;\qy \right)$ based on (\ref{eq:GAUMutuall}) with
 $I\left( \qx ;\qz \right)$ as per (\ref{eq:I_pair}).  Set counter to $n = 1$.

\item Update ${\mathbf{\Lambda}}_{{s}}^{(n)}$ for $s= 1,\ldots,S$ along
the gradient descent direction given by (\ref{eq:gamma_gradient}).

\item Normalize $ \sum\nolimits_{s=1}^{S} {} \big[{\mathbf{\Lambda}}_{{s}}^{(n)} \big]^2 = P$.

\item Update $\mathbf{V}_{{s}}^{(n)}$ for $s= 1,\ldots,S$ along the gradient descent direction in (\ref{eq:V_gradient}).

\item Update $\qXi$, ${\bf{R}}$,
 $\boldsymbol{\gamma}$, and $\boldsymbol{\psi}$
based on (\ref{eq:eqChxi})--(\ref{eq:Varsigma_k-MSE_2}), (\ref{eq:Omega_pair}).

\item Compute $I^{(n + 1)}( \qx;\qy )$ based on (\ref{eq:GAUMutuall}) and (\ref{eq:I_pair}).
If $I^{(n + 1)}( \qx;\qy )  - I^{(n )}( \qx;\qy ) > \varepsilon$
and $n \leq N_{\rm iter}^{\rm max}$,  set $n = n + 1$ and repeat Steps $3$--$7$.

\item Compute ${{{\bf{U}}_{{{\mathrm{\bf{B}}}}}}}$ from the eigenvalue decomposition of the final $\qXi$.

\item Compute ${{{\bf{\Lambda }}_{{{\mathrm{\bf{B}}}}}}}$ and
 ${{{\bf{V}}_{{{\mathrm{\bf{B}}}}}}}$ based on (\ref{eq:Lambda_pair}) and (\ref{eq:V_pair}).  Set $\qB = \mathbf{U}_{{\rm \qB}} {{{\bf{\Lambda }}_{{{\mathrm{\bf{B}}}}}}}
{{{\bf{V}}_{{{\mathrm{\bf{B}}}}}}}$.

 \vspace*{1mm} \hrule

  \end{enumerate}

\end{alg}
\null
\par

\noindent compare the computational complexity of these latter operations.
When $N_{\mathrm t}$ increases, such complexity for the complete-search design in \cite{zeng2012linear}
 scales with $M^{2 N_{\mathrm t} }$. In contrast, for Algorithm \ref{Gradient_Pair}
it scales with $ S  M^{2 N_{\mathrm s} }$.

To illustrate how enormous the savings can be, consider an example where $N_{\mathrm s} = 2$ and the signals are QPSK.
The numbers of additions required by the complete-search design and by Algorithm \ref{Gradient_Pair} are contrasted in Table \ref{tab:mac_dim} for different values of $N_{\mathrm t}$.

\textit{Remark 2:} Through $S$ and $N_{\mathrm s}$, Algorithm \ref{Gradient_Pair} offers a tradeoff between performance and complexity.
At one end, for $S =1$ and $N_{\mathrm s} = N_{\mathrm t}$, Algorithm \ref{Gradient_Pair} searches the entire space while, at the other end,
for $S = N_{\mathrm t}$ and $N_{\mathrm s} = 1$, it merely allocates power among the $N_{\mathrm t}$ parallel subchannels.
Varying $N_{\mathrm s}$ from $1$ to $N_{\mathrm t}$
bridges the gap between separate and fully joint transmission of the $N_{\mathrm t}$ original signals.

%is dominated by the number of additions in calculating the mutual information and the MMSE matrix in  \cite[Eq. (14)]{zeng2012linear}, \cite[Eq. (28)]{zeng2012linear}, and \cite[Eq. (47)]{zeng2012linear}.
 %Accordingly, the  computational complexity of the complete-search design

%As a result, the computational complexity of Algorithm  \ref{Gradient_Pair}  can be significantly lower than that of the complete-search design when the number of transmit antenna is large.

\begin{table}[!t]
\centering
% \captionstyle{center}
\caption{Number of additions required to calculate the mutual information and the MMSE matrix for QPSK and $N_{\rm s} = 2$.} \label{tab:mac_dim}
\vspace*{1.5mm}
\begin{tabular}{|c|c|c|c|c|c|c|c|}
\hline
 $N_{\mathrm t} $   & 4 &  8 &  16 &   32     \\ \hline
  Complete-search  & \multirow{2}{*}{$65536$} & \multirow{2}{*}{$4.29 \times 10^9$}  & \multirow{2}{*}{$1.84 \times 10^{19}$}  &   \multirow{2}{*}{$3.4 \times 10^{38}$}      \\
 design in \cite{zeng2012linear} & & & & \\  \hline
 Algorithm \ref{Gradient_Pair} &   $512$   &   $1024$  & $2048$ & $4096$   \\ \hline
\end{tabular}
\end{table}

% We observe from Table \ref{tab:mac_dim} that Algorithm \ref{Gradient_Pair} requires a significantly
%lower number of  additions  for the MIMO precoder design with finite alphabet input signals and statistical CSI
% compared to the design in \cite{zeng2012linear}.

\textit{Remark 3:}
An adequate choice of $\ell_{1},\ldots, \ell_{N_{\mathrm t}}$ is important for Algorithm \ref{Gradient_Pair}
to perform satisfactorily. The $N_{\mathrm s}/2$ largest diagonal entries of $[\qXi_{\rm eq}]$ should be paired with the $N_{\mathrm s}/2$ smallest diagonal entries.
Then, the $N_{\mathrm s}/2$ next largest diagonal entries of $[\qXi_{\rm eq}]$ should be paired with
the $N_{\mathrm s}/2$ next smallest ones, and so on.
%and  follow the same rule to complete the selection of $\ell_{1}, \ell_{2},\ldots,\ell_{N_{\mathrm t}}$.
%Numerical results indicate that this paring method achieves a good performance with a significantly lower complexity.

\textit{Remark 4:} Since ${\mathbf{\Lambda}}_{{s}}^{(n)}$ and $\mathbf{V}_{{s}}^{(n)}$
are searched along gradient descent directions, in Step 7 the mutual information $I^{\left(n \right)}\left( \qx;\qy \right)$ is nondecreasing.
%At the same time, $I^{\left(n\right)}\left( \qx;\qy \right)$ is upper-bounded.
Algorithm \ref{Gradient_Pair} thus generates sequences that are nondecreasing and upper-bounded, hence it is convergent.
However, due to the nonconvexity of $I^{\left(n \right)}\left( \qx;\qy \right)$ in ${\mathbf{\Lambda}}_{s}^{(n)}$ and $\mathbf{V}_{s}^{(n)}$, Algorithm \ref{Gradient_Pair} may only find local optima.
As a result, the algorithm is run several times with different
initializations of ${\mathbf{\Lambda}}_{{s}}^{(n)}$ and $\mathbf{V}_{{s}}^{(n)}$ and the precoder that provides the highest
mutual information is retained \cite{Wu2012TVT,Wu2012TWC,Wu2013TCOM}.

In the following, we provide an example
to better illustrate the proposed precoder design based on statistical CSI $\bar\qh$, $\qU_{{\rm R}}$, $\tilde{\qG}$, and $\qU_{{\rm T}}$.
%In the following, we provide an example to illustrate about this.

\begin{example} \label{exp:4_1}
Consider a ${1 \times 4}$ deterministic channel  ${{\bf{h}}}_{\rm d} $
%and transmit signal ${\bf{d}} = [d_1,d_2,d_3,d_4]^T $
with SVD ${{\bf{h}}_{\rm d}= \left[{a,0,0,0} \right] {\bf{U}}_{\rm \qh}^H }$ and ${a = \| {\bf{h}}_{\rm d} \| }$. The corresponding received signal is
\begin{equation}
y
%= {\bar{\bf{h}} \bf{U\Lambda Vd}} + n
= \left[ {a,0,0,0} \right]{\bf{U}}_{\rm \qh}^H  {\mathbf{U}_{\rm \qB}  {\boldsymbol{\Lambda}_{\rm \qB}}  \mathbf{V}_{\rm \qB}} \mathbf{d}   + n. \label{eq:y_41_exp}
\end{equation}
Setting ${\bf{U}}_{\rm \qB} = {{\bf{U}}_{\rm \qh}}$ as in \cite{Mohammed2011TIT,Ketseoglou2015TWC}, we obtain
\begin{align}
y = \left[ {a,0,0,0} \right]  {\boldsymbol{\Lambda}_{\rm \qB}}  \mathbf{V}_{\rm \qB}  {\bf{d}} + n. \label{eq:y_41}
\end{align}
If the  precoder were to mix only signals pairs, i.e., $d_1$ with $d_2$ and $d_3$ with $d_4$, then
\begin{align}
%{\bf{\Lambda }} & = \left[ {\begin{array}{*{20}{c}}
%{{\Lambda _1}}&0&0&0\\
%0&{{\Lambda _2}}&0&0\\
%0&0&{{\Lambda _3}}&0\\
%0&0&0&{{\Lambda _4}}
%\end{array}} \right]  \label{eq:model_41_diag} \\
{\bf{V}}_{\rm \qB} & = \left[ {\begin{array}{*{20}{c}}
{{V_{11}}}&{{V_{12}}}&0&0\\
{{V_{21}}}&{{V_{22}}}&0&0\\
0&0&{{V_{33}}}&{{V_{34}}}\\
0&0&{{V_{43}}}&{{V_{44}}}
\end{array}} \right] \label{eq:model_41_V}
\end{align}
%Substituting (\ref{eq:model_41_diag}) and (\ref{eq:model_41_V}) into (\ref{eq:y_41}),
from which
\begin{align}
y = a{\lambda_1}{V_{11}}{d_1} + a{\lambda_1}{V_{12}}{d_2} + n   \label{eq:y_41_2}
\end{align}
which does not contain $d_3$ and $d_4$. If the entries of ${\bf{d}}$ were BPSK-distributed,
the spectral efficiency of (\ref{eq:y_41_2}) could not exceed $2$ b/s/Hz. However, a $1 \times
4$ channel with BPSK inputs can attain $4$ b/s/Hz and thus the precoding is incurring a significant loss.
%The low-complexity precoder clearly results in a significant rate loss in the $1 \times 4$ channel.
%This deficit arises because, for instantaneous-CSI precoding, the number of subchannels that can be paired is $\min({N_{\mathrm{r}}}, {N_{\mathrm{t}}})$ \cite[Sec. IV]{Mohammed2011TIT}, which implies $N_{\mathrm s} = 1$ for
%the $1 \times 4$ channel. However, for the low-complexity design we must have  $N_{\mathrm s} \geq 2$ to at least pair one strong subchannel with one weak subchannel.
%Thus, the low-complexity design with instantaneous CSI in \cite{Mohammed2011TIT,Ketseoglou2015TWC} cannot be implemented effectively in a $1 \times 4$ deterministic channel via $\bar{\qH}$.

Things are better for fading $\qh$, where the low-complexity precoder relies on $\bar\qh$, $\qU_{{\rm R}}$, $\tilde{\qG}$, and $\qU_{{\rm T}}$, as then $\mathbf{U}_{\rm \qB} = \mathbf{U}_{\qXi}$ which in general does not coincide with $ {{\bf{U}}_{\rm \qh}}$; this ensures that all signals reach the receiver.
To gauge the difference, we randomly generate a $1 \times 4$ fading channel $\qh$
based on (\ref{eq:Spatial_Cov}), wherein $K = 1$, $\qU_{{\rm R}} = \qI_{N_{\mathrm{r}}}$, and $\qU_{{\rm T}}$ is a Fourier matrix. Then, we implement
Algorithm \ref{Gradient_Pair} with $N_{\mathrm s} =2$. The spectral efficiency at ${\rm SNR} = 10$ dB is 2.38 b/s/Hz, which exceeds $2$ b/s/Hz. The
corresponding ${\bf{\Lambda}}_{\qXi}^{1/2}$ is
\begin{align}
{\bf{\Lambda}}_{\qXi}^{1/2} = \left[ {\begin{array}{*{20}{c}}
{0.80}&0&0&0\\
0&{0.14}&0&0\\
0&0&{0.28}&0\\
0&0&0&{0.24}
\end{array}} \right], \label{eq:qXi_41}
\end{align}
which indicates that the equivalent channel matrix $\qXi^{1/2}$ in (\ref{eq:EqScalGAUEach0}) is full-rank. For $N_{\mathrm s} =4$, i.e., with full complexity, the spectral efficiency of Algorithm \ref{Gradient_Pair} is $2.40$ b/s/Hz, indicating that
the low-complexity precoder with $N_{\mathrm s} =2$ is close to optimal.

\subsection{Some Special Cases}
\subsubsection{Kronecker Channel Model}

In the Kronecker model, $\bar\qH = {\bf 0}$ and $\mathbf{G}$ is a rank-one matrix of the form
\begin{equation} \label{eq:GKroModel}
    \mathbf{G} = \boldsymbol{\lambda}_{{\rm r}} \boldsymbol{\lambda}_{{\rm t}}^T
\end{equation}
where $\boldsymbol{\lambda}_{{\rm r}}=[\lambda_{{\rm r},1} \ \lambda_{{\rm r},2} \ldots \lambda_{{\rm r},N_{\mathrm r}}]^T \in{\mathbb R}^{N_{\mathrm r}}$ while $\boldsymbol{\lambda}_{{\rm
t}}=[\lambda_{{\rm t},1} \ \lambda_{{\rm t},2} \ldots \lambda_{{\rm t},N_{\mathrm t}}]^T \in {\mathbb R}^{N_{\mathrm t}}$. In this case, (\ref{eq:Spatial_Cov}) can be equivalently written as
\begin{equation}\label{eq:Spatial_Cov_kron}
\qH = \qA_{{\rm R}}^{1/2}  \qW \,  \qA_{{\rm T}}^{1/2}
\end{equation}
where $\qA_{{\rm R}}  = \qU_{\rm R}  {\rm diag}(\boldsymbol{\lambda}_{{\rm r}}) \qU_{\rm R}^H$
and $ \qA_{{\rm T}}  = \qU_{\rm T}  {\rm diag}(\boldsymbol{\lambda}_{{\rm t}}) \qU_{\rm T}^H$.
Then, (\ref{eq:eqChxi}) and (\ref{eq:eqChMatrixTR}) reduce to
\begin{align} \label{eq:eqChMatrixTR_Kron}
\qXi & =   \mathbf{T} = \gamma^{\circ} \mathbf{A}_{{\rm T}}
\end{align}
and
\begin{align}
\mathbf{R} & = \psi^{\circ} \mathbf{A}_{{\rm R}}
\end{align}
where $\gamma^{\circ} = \boldsymbol{\lambda}_{{\rm r}}^T \boldsymbol{\gamma} $ and $\psi^{\circ} = \boldsymbol{\lambda}_{{\rm t}}^T
\boldsymbol{\psi}$. Thus, from (\ref{eq:Varsigma_k-MSE}),
\begin{align}
\gamma^{\circ}& =\tr \! \left( \left(\mathbf{I}_{N_{\mathrm r}}+\mathbf{R}\right)^{-1} \mathbf{A}_{{\rm r}_{i_k}} \right) \label{eq:Varsigma_k-MSE_MuKron1}\\
\psi^{\circ}& = \tr  ( \boldsymbol{\Omega} \mathbf{A}_{{\rm T}} ). \label{eq:Varsigma_k-MSE_MuKron2}
\end{align}

From (\ref{eq:eqChMatrixTR_Kron}), the optimal left singular matrix $\mathbf{U}_{{\mathrm{\bf B}}}$ of the precoder $\qB$ for this channel model equals
$\mathbf{U}_{{\rm T}}$.
Hence, the equivalent channel matrix between $\qx_{\rm eq}$ and $\qz_{\rm eq}$ in (\ref{eq:EqScalGAUEach}) simplifies to $\sqrt{\gamma^{\circ}} \, \diag(\boldsymbol{\lambda}_{{\rm t}})^{1/2}$.
Also, (\ref{eq:Varsigma_k-MSE_MuKron1}) and (\ref{eq:Varsigma_k-MSE_MuKron2}) indicate that instead of
computing $N_{\mathrm t} + N_{\mathrm r}$ parameters in fix-point equation (\ref{eq:Varsigma_k-MSE}),
we need only compute $\gamma^{\circ}$ and $\psi^{\circ}$ in Algorithm 1. { Furthermore,
the receiver needs to feed back only $\mathbf{A}_{{\rm T}}$ and $\mathbf{A}_{{\rm R}}$ to the transmitter for
precoder design. }

\subsubsection{Deterministic Channel}

For $K\rightarrow\infty$, the random portion of the channel vanishes and
(\ref{eq:GAUMutuall}) becomes
\begin{align}
I_{\rm asy}(\qx;\qy) = I\left( \qx; \bar{\qH} \qx + \check{\bf n} \right)
\label{eq:I_asy}
\end{align}
%where $\qXi$ in (\ref{eq:eqChxi}) reduces to
%\begin{align}
%    \qXi =  \bar{\qH}^H  \bar{\qH}.  \label{eq:eqChxi_2}
%\end{align}
which is exact regardless of the dimensionality.
%From (\ref{eq:I_asy}) and (\ref{eq:eqChxi_2}), we know that the asymptotic approximation in Proposition \ref{Proposition_1} becomes exact mutual information when the channel approaches deterministic.
In this case, { the receiver only needs to feed back $\bar{\qH}$ to the transmitter for precoder design.}

\end{example}

%Then, using Proposition \ref{Proposition_1}, (\ref{eq:EqScalGAUEach}), and (\ref{eq:I_pair}), we  obtain the asymptotic mutual information expression for the
%proposed low complexity design {\rl
%\begin{align}\label{eq:GAUMutuall_MuKron}
% & I_{\rm asy}^{\rm kron }(\qx;\qy)  \nonumber \\
% & \hspace{-0.3cm} \simeq I_{\rm kron }(\qx_{\rm eq};\qz_{\rm eq})  + \log_2 \det\left(\mathbf{I}_{N_{\mathrm r}}+\mathbf{R}\right)  - \log_2 e  \gamma^{\circ} \psi^{\circ}\\
%  & \hspace{-0.3cm} =   \sum\limits_{s = 1}^S I_{\rm kron}\left( {{\bf{d}}_{s}};\qz_{s}  \right) + \log_2 \det\left(\mathbf{I}_{N_{\mathrm r}}+\mathbf{R}\right)  - \log_2 e  \gamma^{\circ} \psi^{\circ}
%\end{align}
%where the  equivalent channel between $\qx_{\rm eq}$ and $\qz_{\rm eq}$ is $ \sqrt{\gamma^{\circ}} {\rm diag}(\boldsymbol{\lambda}_{{\rm t}})^{1/2}$. }

\subsubsection{Massive MIMO}

In some cases, by exploiting the spatial characteristics of physical channels,
the structure of $\bar{\qH}$ can acquire a particular relationship with respect to $\qU_{{\rm T}}$ and $\qU_{{\rm R}}$. Then, Algorithm 1 can be simplified.

Assume there are $L + 1$ independent paths between the transmitter
and the receiver, where the $0$th path is the LOS path. Let $c_l$, $\phi_{l,d}$,
and $\theta_{l,a}$ denote the attenuation, the angle of departure,
and the angle of arrival for the $l$th path. Then, the $N_{\mathrm r} \times N_{\mathrm t}$ MIMO channel
can be modeled as \cite[Sec. 7.3.2]{Tse2005}
\begin{align} \label{eq:H_phy}
{\bf{H}} = {c_0} \, {e^{ - j2\pi {d_0}/{\lambda _c}}} \,{{\bf{u}}_{\mathrm r}} ( {{\theta _{0,a}}} ) \, {\bf{u}}_{\mathrm t}^H ( {{\phi _{0,d}}} ) + \sum\limits_{l = 1}^L {{c_l} \, {e^{ - j2\pi {d_l}/{\lambda_{\rm c}}}} \, {{\bf{u}}_{\mathrm r}} ( {{\theta _{l,a}}} ) \, {\bf{u}}_{\mathrm t}^H  ( {{\phi _{l,d}}} )}
\end{align}
where $d_l$ denotes the distance between transmit antenna 1 and receive antenna 1 along path $l$ and
$\lambda _{\rm c}$ denotes the wavelength;
${\bf{u}}_{\mathrm t}(\phi)\in\mathbb{C}^{N_{\mathrm t}\times 1}$ and ${\bf{u}}_{\mathrm r}(\theta)\in\mathbb{C}^{N_{\mathrm r}\times 1}$ are the unit-norm transmit and receive array response vectors.
%corresponding to the angle of departure (AoD) satisfying ${\bf{u}}^H_{\mathrm t}(\phi){\bf{u}}_{\mathrm t}(\phi) = 1$.
%corresponding to the angle of arrival (AoA) satisfying ${\bf{u}}^H_{\mathrm r}(\theta){\bf{u}}_{\mathrm r}(\theta) = 1$.

%For the LOS path, ${c_0}{e^{ - j2\pi {d_0}/{\lambda _c}}}$ is deterministic.
%For the fading path $l$, we have $E \left[{c_l}{e^{ - j2\pi {d_l}/{\lambda _c}}}
% \right] = 0$, $l = 1,2,\ldots,L$. In addition, we assume different paths are mutually independent.
In massive MIMO \cite{5G},
 the array response vectors become asymptotically orthogonal\cite{Viberg1995TSP,Rusek2013SPM}, i.e.,
%\footnote{Massive antenna arrays can provide high resolution of the signal angles.
%Therefore, arrays with an infinite number of antennas can distinguish signals
%from different angles even with infinitesimal spatial separation \cite{}.}
\begin{align}\label{eq:AoD}
\lim_{N_{\mathrm t} \rightarrow \infty}{\bf{u}}_{\mathrm t}^H ( \phi_p  ) \, {{\bf{u}}_{\mathrm t}} ( \phi_l  ) &= \delta ( {p  - l } )
\end{align}
where $\delta ( {p  - l } )$ denotes the Dirac delta pulse.
Under this condition,
%For the user side, when the sampling of $\theta_p$ satisfies
%the condition
%\footnote{The angle $\theta_p$ is the sampling of receive signals. When the response
%vectors are orthogonal, the user can separate these orthogonal direction signals
%perfectly. For ULA, uniform sampling of $\sin(\theta_p)$,
%i.e., $\sin(\theta_p) = n/N_{\mathrm r}$, is a classical choice for spatial angles \cite{sayeed2002deconstructing}.}
%\begin{align}\label{eq:AoA}
%{\bf{u}}_{\mathrm r}^H\left( \theta_p  \right){{\bf{u}}_{\mathrm r}}\left( \theta_l  \right) &= \delta \left( {p  - l } \right)
%\end{align}
the channel matrix in (\ref{eq:H_phy}) can be rewritten as
\begin{align}\label{eq:H}
{\bf{H}} & = \sum\limits_{n = 1}^{{N_{\mathrm r}}} {\sum\limits_{m = 1}^{{N_{\mathrm t}}} {{{\left[ {\tilde {\bf{H}} + \hat{{\bf{H}}}} \right]}_{nm}}{{\bf{u}}_{\mathrm r}} ( {{\theta _n}} )} \, {\bf{u}}_{\mathrm t}^H} ( {{\phi _m}} ) \\
& =  \qU_{{\rm R}} \left( {\tilde {\bf{H}} +  \hat{\bf{H}}}  \right)  \qU_{{\rm T}}^H \label{eq:H_2}
\end{align}
where $\qU_{{\rm T}} = \left[\qu_{\mathrm t}\left(\phi_1\right),\qu_{\mathrm t}\left(\phi_2\right),\ldots,\qu_{\mathrm t} \left(\phi_{N_{\mathrm t}}\right)\right]$
and $\qU_{{\rm R}} = \left[\qu_{\mathrm r}\left(\theta_1 \right),\qu_{\mathrm r}\left(\theta_2 \right),\ldots,\qu_{\mathrm r} \left(\theta_{N_{\mathrm r}}\right)\right]$
are unitary. Then,
the entries of ${\widetilde {\bf{H}}}$ and $\hat{\bf{H}}$ satisfy \cite{sayeed2002deconstructing}
\begin{align}
{\left[ {\tilde {\bf{H}}} \right]_{nm}} &\simeq \sum\limits_{l \in {F_{r,n}} \cap {F_{t,m}}} \!\! {{c_l} \, {e^{ - j2\pi {d_l}/{\lambda _{\rm c}}}}} \label{eq:H_fading} \\
{\left[ {\hat{{\bf{H}}} } \right]_{nm}} &\simeq \left\{ \begin{array}{lll}
{c_0} \, {e^{ - j2\pi {d_0}/{\lambda _{\rm c}}}} &{}&\quad {T(n,m) = 1} \\
0 &{}&\quad  {\rm otherwise}.
\end{array} \right. \label{eq:H_deterministic}
\end{align}
where ${F_{r,n}}$ and ${F_{t,m}}$ denote the subsets of paths whose
angles are closest to $\theta_n$ and $\phi_m$, respectively. In turn, $T(n,m) = 1$ if the angles of the LOS path
are closest to $\theta_n$ and $\phi_m$ simultaneously; for other $n$ and $m$, conversely, $T(n,m) = 0$.
It should be noted that the approximations in (\ref{eq:H_fading}) and
(\ref{eq:H_deterministic}) become exact
when the dimension of the antenna tends to infinity \cite{Sun2015TCOM}.

Stacking the columns of $\qH$ into a vector, we obtain
\begin{equation}\label{eq:H_p_vector}
{\rm vec} ( {{{\bf{H}}}} ) = \sum\limits_{n = 1}^{{N_{\mathrm r}}} {\sum\limits_{m = 1}^{{N_{\mathrm t}}} {\left( {{{\big[ {{\bf{\tilde H}}} \big]}_{nm}} + {{\big[ {\hat {\bf{H}}} \big]}_{nm}}} \right)
\left({\bf{u}}_{\mathrm t}^* ( {{\phi _m}} ) \otimes {{\bf{u}}_{\mathrm r}}( {{\theta _n}} )\right)} }
\end{equation}
from which the correlations within $\bf H$ are completely characterized as
\begin{align}\label{eq:full_cov}
&  E \! \left[{\rm vec}( {{{\bf{H}}}} ) {\rm vec}( {{{\bf{H}}}} )^H \right]  = \sum\limits_{n = 1}^{{N_{\mathrm r}}}  \sum\limits_{m = 1}^{{N_{\mathrm t}}}  E \! \left[  {{{\big[ {{\bf{\tilde H}}} \big]}_{nm}}    {{\big[ {{{{\bf{\tilde H}}}^H}} \big]}_{nm}}}  \right]  \Big( {{\bf{u}}_{\mathrm t}^*( {{\phi _m}} ) \otimes {{\bf{u}}_{\mathrm r}}( {{\theta _n}} )} \Big)       {\Big( {{\bf{u}}_{\mathrm t}^*( {{\phi _m}} ) \otimes {{\bf{u}}_{\mathrm r}}( {{\theta _n}} )} \Big)^H}  \nonumber \\
& \qquad\qquad +  \sum\limits_{n = 1}^{{N_{\mathrm r}}}  \sum\limits_{m = 1}^{{N_{\mathrm t}}} \left( {{\big[ {\hat {\bf{H}}} \big]}_{nm}} {{\big[ {\hat {\bf{H}}^H} \big]}_{nm} } \right)\Big( {{\bf{u}}_{\mathrm t}^*( {{\phi _m}} ) \otimes  {{\bf{u}}_{\mathrm r}} ( {{\theta _n}} )} \Big)   {\Big( {{\bf{u}}_{\mathrm t}^* ( {{\phi _m}} )  \otimes {{\bf{u}}_{\mathrm r}} ( {{\theta _n}} )} \Big)^H}.
\end{align}
The first term on the right-side of (\ref{eq:full_cov}) equals the correlation matrix of the first term on the right-side of (\ref{eq:Spatial_Cov}).
Thus,
\begin{equation}\label{eq:G_H}
{\bf{G}} = E\left[ {{\bf{\tilde H}} \odot {\bf{\tilde H}}^{*}} \right] .
\end{equation}
%where the $(n,m)$th element of $\qG$ represents the average amount of energy
%that is coupled from $\qu_{{\rm R},n}$ to $\qu_{{\rm T},m}$.
Comparing (\ref{eq:Spatial_Cov}) and (\ref{eq:H_2}), we have that
\begin{align}
\label{eq:sta_phy}
{\bf{\tilde H}} & = {\bf{\tilde G}} \odot {\bf{W}} \\
{\bf{\bar H}} & = \qU_{{\rm R}}  \hat{\bf{H}}  \qU_{{\rm T}}^H
\label{eq:sta_phy2}
\end{align}
which relate the massive MIMO channel with the model used in our analysis.
The sum of fading paths in (\ref{eq:H_fading}) can be modeled as a Gaussian random
variable with variance  $[\qG]_{ij}$ while the LOS path can be modeled as a rank-one matrix having the same
transmit and receive eigendirections as the fading paths, i.e., $\qU_{{\rm T}}$ and $\qU_{{\rm R}}$.

%By substituting  ${\bf{\bar H}} = \qU_{{\rm R}}  \hat{\bf{H}}  \qU_{{\rm T}}^H$ into (\ref{eq:eqChxi}), we know
From Proposition \ref{prop:opt_structure}, the optimal $\qU_{\rm \qB}$ equals $\qU_{{\rm T}}$. Plugging such matrix
into (\ref{eq:x}), using \cite[(5)]{zeng2012linear} and
recalling (\ref{eq:sta_phy}), (\ref{eq:sta_phy2}), we can re-write  (\ref{eq:x}) as
\begin{equation}\label{eq:y_eq_phy}
{{\bf{y}}_{{\rm{phy}}}} = {{\bf{H}}_{{\rm{phy}}}} \qx_{\rm phy}  + {\bf{{n}}}
\end{equation}
where
\begin{align}\label{eq:x_eq_phy}
\qx_{\rm phy} &=  { {{\bf{\Lambda }}_{{{\rm{\qB}}}}}{{\bf{V}}_{{{\rm{\qB}}}}} } \qd \\
\label{eq:H_eq_phy}
{{\bf{H}}_{{\rm{phy}}}}& = \big( \tilde{\qG} \odot \qW \big)  +  \hat{\bf{H}}.
\end{align}
With that, (\ref{eq:eqChxi}) becomes
\begin{align}
    \qXi_{\rm phy} & = \qT_{\rm phy} + \hat{\qH}^H \left(\qI_{N_{\mathrm r}}+ \qR_{\rm phy}\right)^{-1} \hat{\qH} \label{eq:eqChxi_eq_phy}
\end{align}
where
\begin{align} \label{eq:eqChMatrixTR_2_phy}
\qT_{\rm phy} & =  {\diag}( \qG^T \boldsymbol{\gamma}_{\rm phy} ) \\
\qR_{\rm phy} & =  {\diag} ( \qG \boldsymbol{\psi}_{\rm phy} ).
\end{align}
The entries of $\boldsymbol{\gamma}_{\rm phy}$ and $\boldsymbol{\psi}_{\rm phy}$
are the solution to the fixed point-equations
\begin{align}
 [\boldsymbol{\gamma}_{\rm phy} ]_{m} & = \left[  \left(\qI_{N_{\mathrm r}}  +  \qR_{\rm phy}\right)^{-1} \left( \qI_{N_{\mathrm r}}  -  \hat{\qH} \qOmega_{\rm phy} \hat{\qH}^{H} \left(\qI_{N_{\mathrm r}}
 +  \qR_{\rm phy}\right)^{-1} \right)  \right]_{mm} \nonumber \\
[\boldsymbol{\psi}_{\rm phy} ]_{n} & =  [ \qOmega_{\rm phy} ]_{nn} \label{eq:Varsigma_k-MSE_2_phy_2}
\end{align}
where
\begin{align}\label{eq:Omega_eq}
    \qOmega_{\rm phy} & = E \! \left[ (\qx_{\rm phy} - \hat\qx_{\rm phy} ) (\qx_{\rm phy} - \hat\qx_{\rm phy})^H \right]  \\
   \hat\qx_{\rm phy} &= E [\qx_{\rm phy} |\qz ].
 \end{align}

In massive MIMO, altogether, Algorithm \ref{Gradient_Pair}
can be  simplified in two ways. First, Step 8 is rendered unnecessary since $\mathbf{U}_{{\rm \qB}} = \mathbf{U}_{{\rm T}}$.
Second, in Steps 2 and 6
the fixed-point equations (\ref{eq:eqChxi_eq_phy})--(\ref{eq:Varsigma_k-MSE_2_phy_2}) involve only diagonal matrices, with the ensuing computational simplification. { Furthermore, the receiver needs to feed back the non-zero elements of ${\hat{{\bf{H}}} }$
in (\ref{eq:H_deterministic}), ${\bf{\tilde G}}$, $\qU_{{\rm R}}$, and $\qU_{{\rm T}}$ to the transmitter for precoder design.}
% can be implemented more easily in practical systems than that of computing the original fix point equations in (\ref{eq:eqChxi})--(\ref{eq:Varsigma_k-MSE}).

\section{Performance Evaluation}

%For the purpose of precoder comparisons, let us denote by GP, NP and FAP the Gaussian precoding, no precoding,
%and finite alphabet precoding.

First, let us evaluate the complexity of Algorithm 1 for different values of $N_{\mathrm s}$.
%We evaluate the running time for performing one iteration in Algorithm 1 with different antennas numbers, different $N_{\mathrm s}$, and different modulations.
{ Tables II--IV provide the number of additions required to calculate the mutual information and the MMSE matrix per iteration of
Algorithm 1 for various numbers of antennas and different signal constellations.}
%The evaluation results are given in Table I--Table III,
As anticipated, for $N_{\mathrm s} = N_{\mathrm t}$, the computational complexity grows exponentially with $N_{\mathrm t}$ and quickly
becomes unmanageable.
{
\begin{table}[!t]

\label{runing_time_1}
\centering
 \captionstyle{center}
  {
\caption{Number of additions required to calculate the mutual information and the MMSE matrix with BPSK}
\vspace{5pt}
\begin{tabular}{|c|c|c|c|}
\hline
  $N_{\mathrm t}$  &     $ N_{\mathrm s} = 2$   &   $ N_{\mathrm s} = 4$  &  $N_{\mathrm s} = N_{\mathrm t}$        \\ \hline
  $4$  & 32   &   256   &  256   \\ \hline
 $8$    &  64   &  512  &   65536     \\ \hline
%  $12$    &  0.0172   &   0.0762 &  $\times$       \\ \hline
   $16$    & 128   &  1024  &   4.2950e+009     \\ \hline
      $32$    &  256    &  2048 &   1.8447e+019     \\ \hline
\end{tabular}
}
\end{table}

\begin{table}[!t]

\label{runing_time_2}
\centering
\captionstyle{center}
  {
\caption{Number of additions required to calculate the mutual information and the MMSE matrix with QPSK}
\vspace{5pt}
\begin{tabular}{|c|c|c|c|}
\hline
  $N_{\mathrm t}$  &     $ N_{\mathrm s} = 2$   &   $ N_{\mathrm s} = 4$  &  $N_{\mathrm s} = N_{\mathrm t}$        \\ \hline
  $4$  &  512  &    65536   &   65536 \\ \hline
 $8$    & 1024  &   131072  &   4.2950e+009       \\ \hline
%  $12$    &  0.2178   &   36.8507  &  $\times$       \\ \hline
   $16$    & 2048   &   262144 &   1.8447e+019     \\ \hline
      $32$    & 4096  &  524288 &     3.4028e+038    \\ \hline
\end{tabular}
}
\end{table}

\begin{table}[!t]

\label{runing_time_3}
\centering
 \captionstyle{center}
  {
\caption{Number of additions required to calculate the mutual information and the MMSE matrix with 16-QAM.}
\vspace{5pt}
\begin{tabular}{|c|c|c|}
\hline
  $N_{\mathrm t}$  &     $ N_{\mathrm s} = 2$    &  $N_{\mathrm s} = N_{\mathrm t}$        \\ \hline
  $4$  &  512   &   4.2950e+009  \\ \hline
 $8$    &  1024 &     1.8447e+019        \\ \hline
 % $12$    &  89.0969  &  $\times$       \\ \hline
   $16$    &  2048     &  3.4028e+038     \\ \hline
      $32$    & 4096 &   1.1579e+077     \\ \hline
\end{tabular}
}
\end{table}
}

{ Fig. \ref{QPSK_Rician_asy} compares the spectral efficiency vs.
Rice factor $K$ for the channel in (\ref{eq:Spatial_Cov}) with $N_{\mathrm t} = N_{\mathrm r} =4$, ${\rm SNR} = 15$ dB, and QPSK.
$\mathbf{U}_{\rm R}$, $\mathbf{U}_{\rm T}$, and $\mathbf{\tilde{G}}$ in (\ref{eq:Spatial_Cov}) are generated randomly.
The Rice component in (\ref{eq:Spatial_Cov}) is generated based on the physical channel model in (\ref{eq:H_phy}). As illustrated in Fig. 1, even for a small number of antennas,
the spectral efficiency of the proposed low complexity design with $N_{\mathrm s} = 2$ is close to the spectral efficiency
of the complete search design with $N_{\mathrm s} = 4$ for a large range of Rician factors $K$.
Also, the approximated spectral efficiency  in (\ref{eq:GAUMutuall}), denoted by ``Asymptotic" in Fig. 1,
is close to the exact spectral efficiency in (\ref{eq:Mutual_Info}). {\bl The exact expression in (14) is computed via a Monte Carlo average over the channel matrix $\mathbf{H}$.}

\begin{figure}[!t]
\centering
\includegraphics[width=0.8\textwidth]{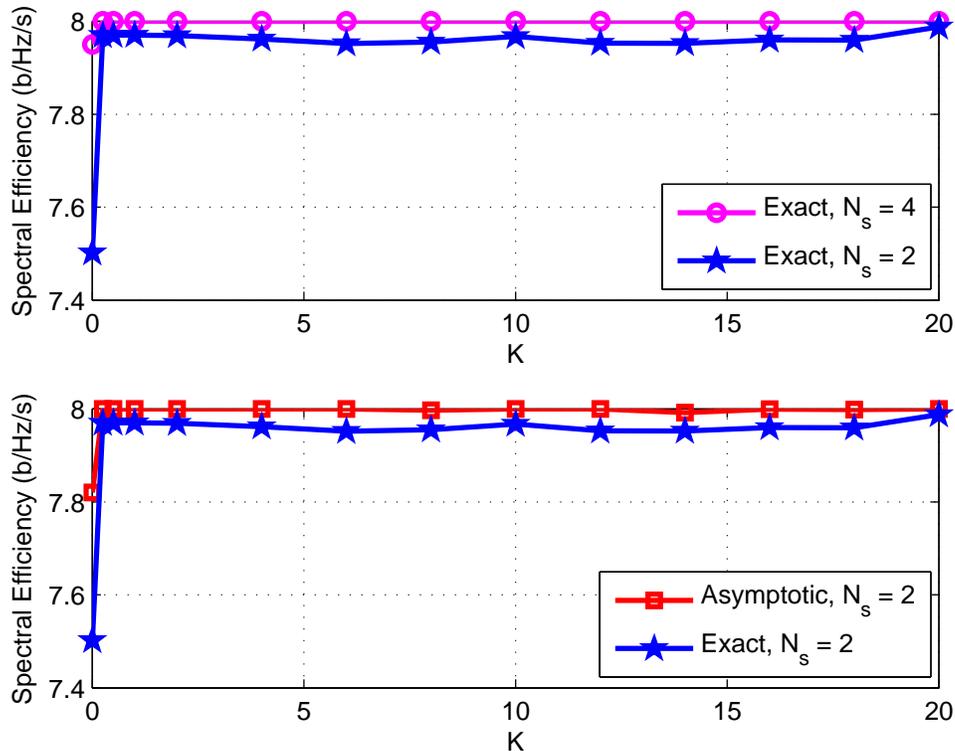}
 \captionstyle{flushleft}
 \vspace{-10pt}
 \caption{Spectral efficiency vs. Rice factor $K$ for the  channel in (\ref{eq:Spatial_Cov}) with $N_{\mathrm t} = N_{\mathrm r} =4$,
 ${\rm SNR} = 15$ dB, and QPSK.}
\label{QPSK_Rician_asy}
\end{figure}
}

{Next, we examine Algorithm \ref{Gradient_Pair} for practical channels.
We adopt the 3GPP SCM \cite{Salo2005} for the urban scenario, half-wavelengh antenna spacing at transmitter and receiver, respectively,
a velocity of $36$ km/h, and $6$ paths. We obtain
$\bar\qH$, $\qU_{{\rm R}}$, $\tilde{\qG}$, and $\qU_{{\rm T}}$ based on a large number of  these realizations
for the SCM model, and use them for precoder design. }

\begin{figure}[!t]
\centering
\includegraphics[width=0.8\textwidth]{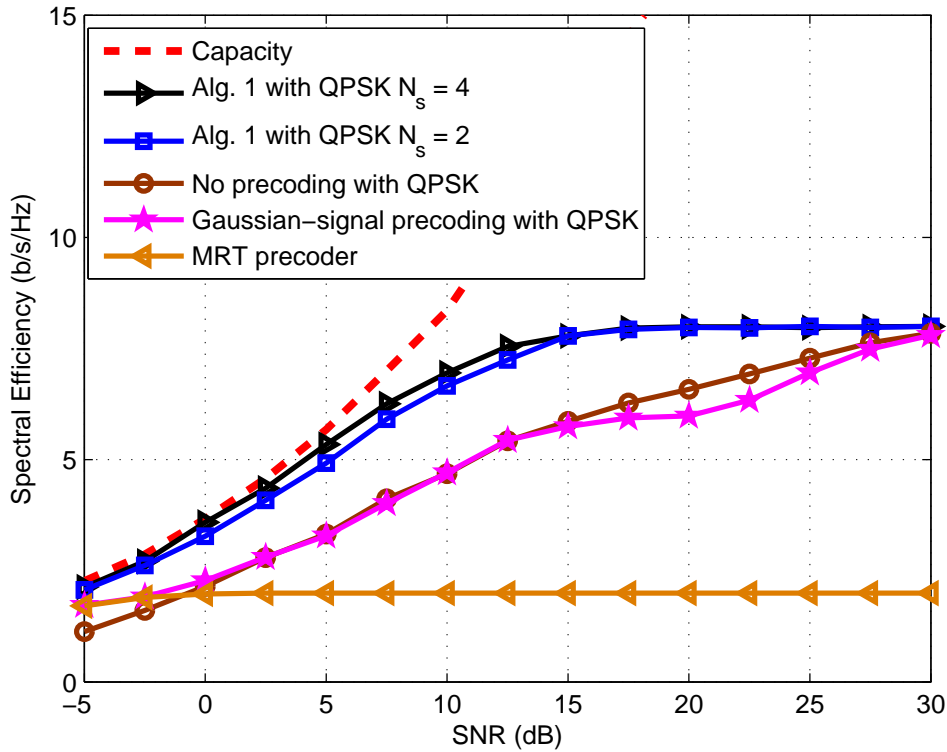}
 \captionstyle{flushleft}
 \vspace{-10pt}
 \caption{Spectral efficiency vs. SNR for the 3GPP SCM (urban scenario, half-wavelength antenna spacing, $36$ km/h) for different precoder designs with $N_{\mathrm t} = N_{\mathrm r} =4$ and QPSK.}
\label{QPSK_SCM}
\end{figure}

Fig. \ref{QPSK_SCM} depicts the spectral efficiency for the 3GPP SCM
for different precoder designs with $N_{\mathrm t} = N_{\mathrm r} = 4$ and QPSK.
%Massive channel realizations were generated based on SCM and statistical CSI in (\ref{eq:Spatial_Cov}) can be acquired from these channel realizations.
A Gauss-Seidel algorithm with stochastic programming is employed
to obtain the capacity-achieving precoder \cite{wen2011sum}.
{\bl Also, the performance of the
maximum ratio transmission precoder from \cite{Lo1999TCOM} is simulated, denoted by ``MRT precoder".}
We substitute the final precoder matrices obtained
by different designs into (\ref{eq:Mutual_Info}) to evaluate the ergodic spectral efficiency.
For Algorithm \ref{Gradient_Pair}, both $N_{\mathrm s} = 4$ and $N_{\mathrm s} = 2$ are considered, and despite their enormous computational gap (cf. Table IV) the difference in performance is minor.
Both precoders hug the capacity up to the point where the QPSK cardinality becomes insufficient.
%which is similar to the spectral efficiency of precoder designed in \cite{zeng2012linear}.?The precoder designed via Algorithm \ref{Gradient_Pair}
{\bl The proposed design gains many dB over an unprecoded transmitter, the capacity-achieving precoder applied with QPSK, and
the MRT precoder. It is observed in Fig. 2 that, when SNR is low, the performance of the MRT and the capacity-achieving precoders is close to that of
the proposed design. This is because the MRT precoder is actually a beamformer and, in the low SNR regime, the beamforming design is near-optimal
for  both  Gaussian input and finite-alphabet inputs \cite{Perez-Cruz2010TIT}. However, as the SNR increases, the beamforming design
results in a pronounced performance loss, as shown in Fig. 2. }

%We obtain the precoders via Algorithm \ref{Gradient_Pair} and then substitute them into (\ref{eq:Mutual_Info}) to calculate the exact finite alphabet rate.
%From Figure \ref{QPSK_SCM}, we observe that the mutual information achieved by the ``FAP with QPSK inputs $N_{\mathrm s} =4$" design and the ``FAP with QPSK inputs $N_{\mathrm s} =2$" design are almost the same.
%However, as indicated in Table II, the computation complexity of the ``FAP with QPSK inputs $N_{\mathrm s} =4$" design is hundreds of times higher than that of the ``FAP with QPSK inputs $N_{\mathrm s} =2$" design.
%Also, we note from Figure \ref{QPSK_SCM}  that both the ``FAP with QPSK inputs $N_{\mathrm s} =4$"
%and the ``FAP with QPSK inputs $N_{\mathrm s} =2$" designs perform closely to the channel capacity
%in the low and moderate SNR regions and saturate at $N_{\mathrm t} \log_2 M = 8$ b/Hz/s at high SNR.
%In addition, both designs  achieve substantial SNR gains over
%the ``NP with QPSK inputs" design and the ``GP with QPSK inputs" design.

\begin{figure}[!t]
\centering
\includegraphics[width=0.8\textwidth]{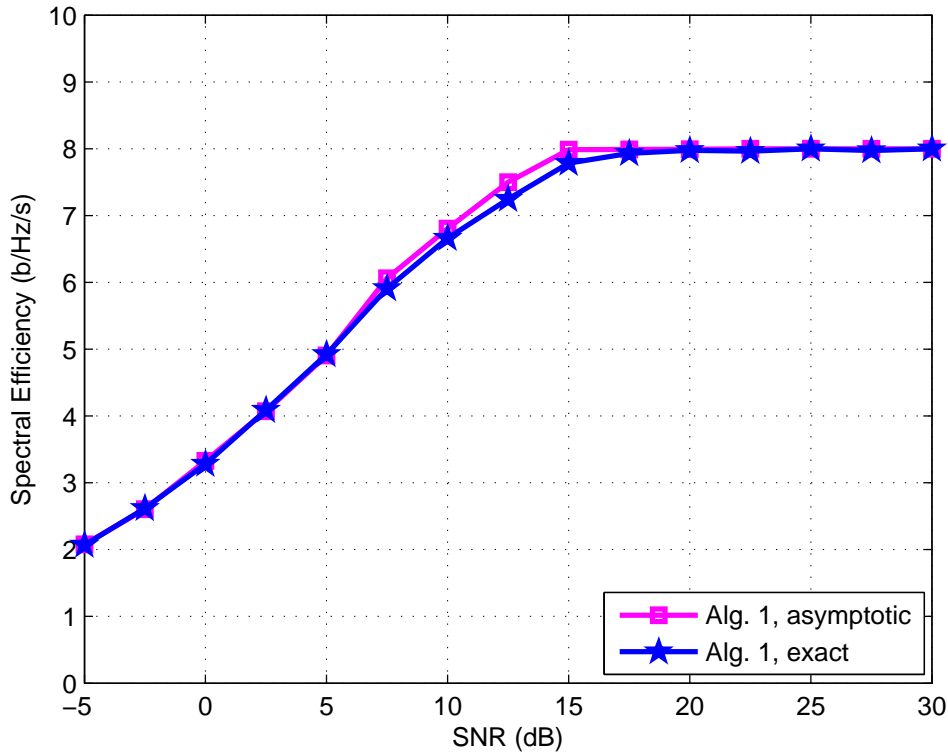}
 \captionstyle{flushleft}
  \vspace{-10pt}
\caption{Asymptotic and exact spectral efficiency vs. SNR for the 3GPP SCM (urban scenario, half-wavelength antenna spacing, $36$ km/h) with $N_{\mathrm t} = N_{\mathrm r} =4$, $N_{\mathrm s} =2$, and QPSK.}
\label{QPSK_SCM_asy}
\end{figure}

Fig. \ref{QPSK_SCM_asy} contrasts the spectral efficiency given by the asymptotic expression in (\ref{eq:GAUMutuall}) with the exact form in (\ref{eq:Mutual_Info}) for the precoders obtained
by Algorithm \ref{Gradient_Pair} with $N_{\mathrm s} = 2$.  The channel model is the same as for Fig. \ref{QPSK_SCM}.
We observe from Fig. \ref{QPSK_SCM_asy} that even for a small number of antennas,
the asymptotic spectral efficiency  in (\ref{eq:GAUMutuall})  is close to the exact spectral efficiency.

\begin{figure}[!t]
\centering
\includegraphics[width=0.8\textwidth]{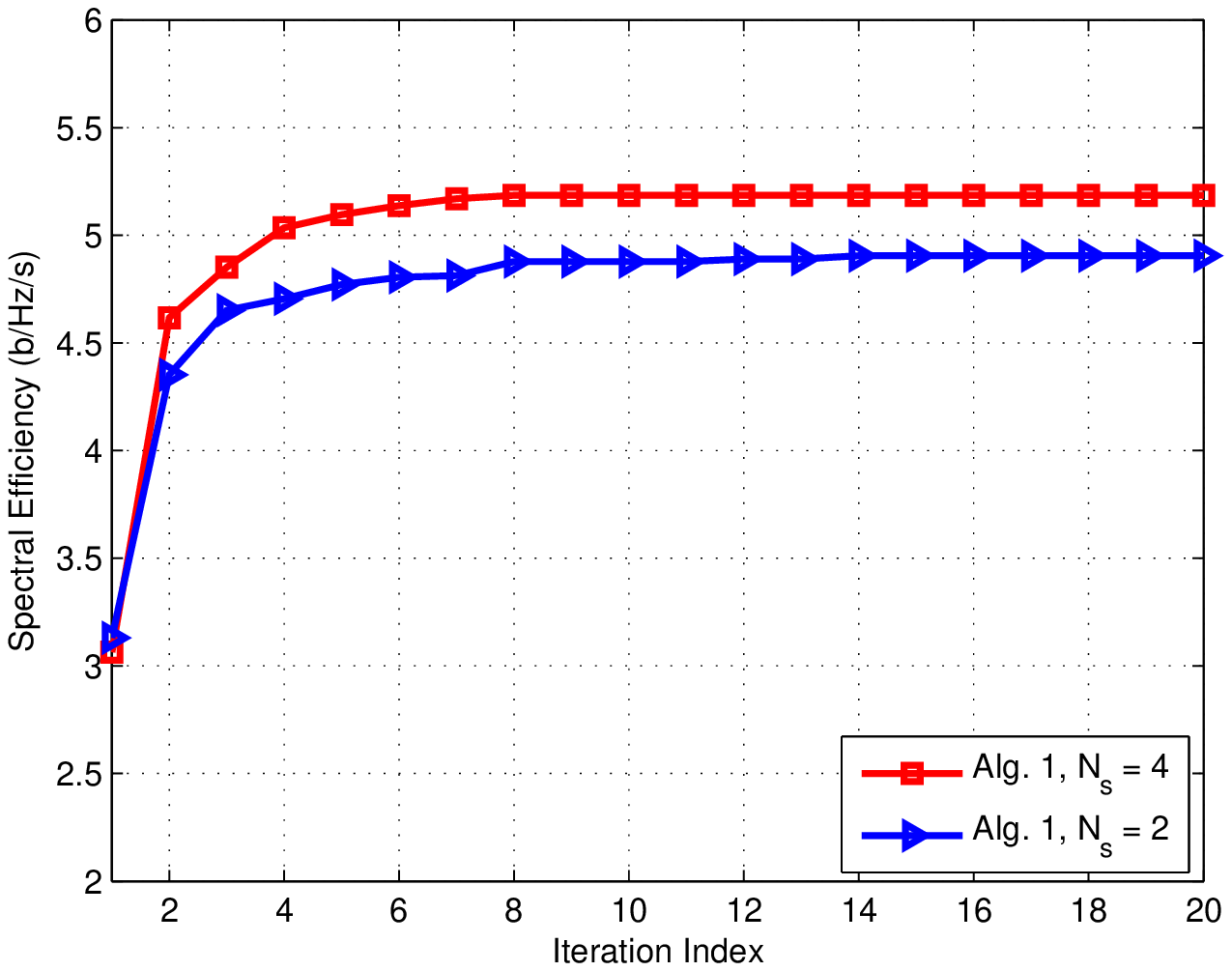}
 \captionstyle{flushleft}
  \vspace{-10pt}
\caption{Convergence of Algorithm \ref{Gradient_Pair} for the 3GPP SCM (urban scenario, half-wavelength antenna spacing, $36$ km/h) with $N_{\mathrm t} = N_{\mathrm r} =4$ and QPSK.}
\label{QPSK_SCM_converge}
\end{figure}

Fig. \ref{QPSK_SCM_converge} illustrates the rapid convergence of Algorithm \ref{Gradient_Pair} for
$N_{\mathrm s} =2$ and $N_{\mathrm s} = 4$ at ${\rm SNR} = 5$ dB. The channel model is the same as for Fig. \ref{QPSK_SCM}.
%We can see from Figure \ref{QPSK_SCM_converge} that  Algorithm \ref{Gradient_Pair} converges in a few iterations.

\begin{figure}[!t]
\centering
\includegraphics[width=0.8\textwidth]{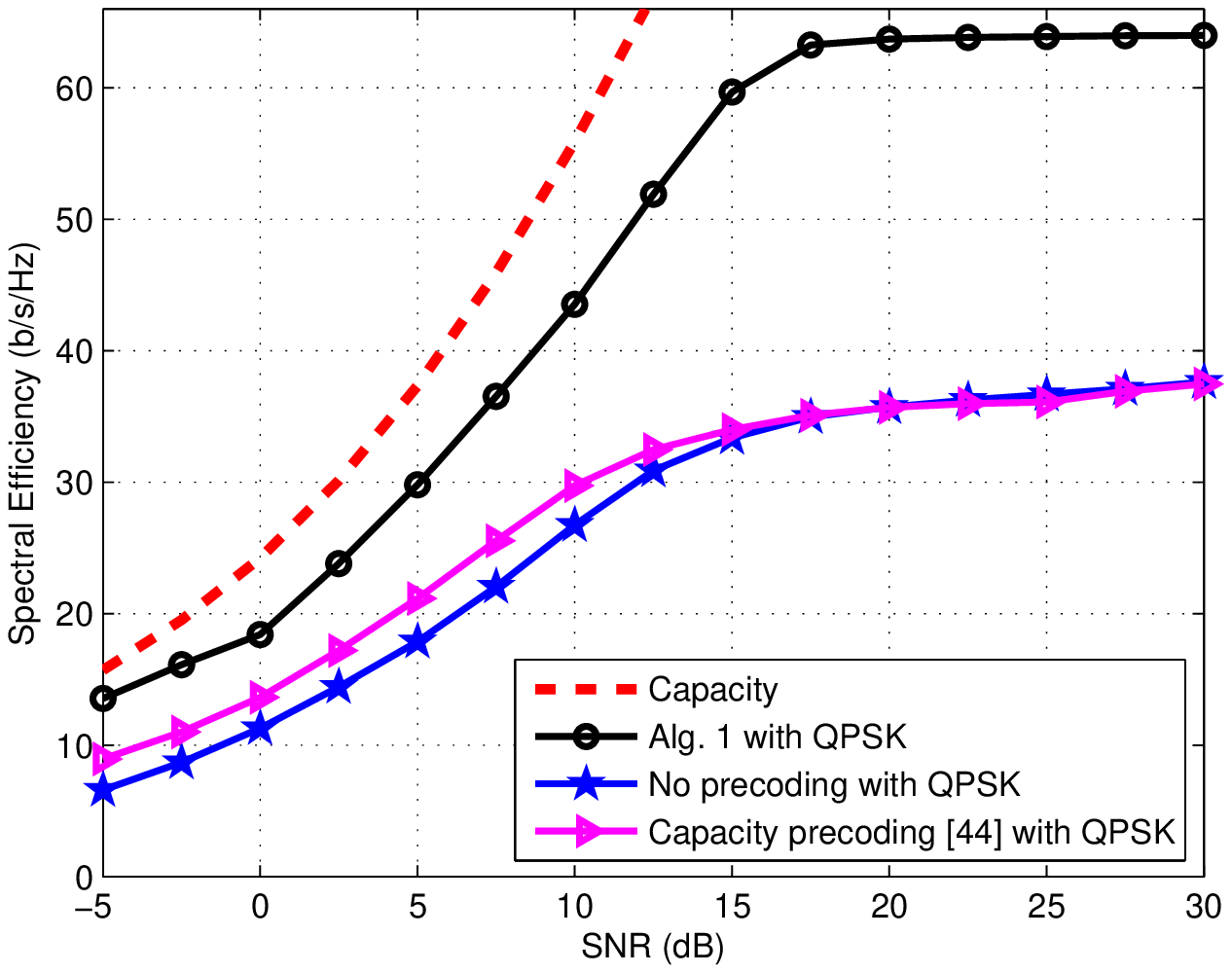}
 \captionstyle{flushleft}
  \vspace{-10pt}
\caption{Spectral efficiency vs. SNR for the 3GPP SCM (urban scenario, half-wavelength antenna spacing, $36$ km/h) for different precoder designs with $N_{\mathrm t} = N_{\mathrm r} =32$, $N_{\mathrm s} = 4$, and QPSK.}
\label{QPSK_SCM_32}
\end{figure}

\begin{figure}[!t]
\centering
\includegraphics[width=0.8\textwidth]{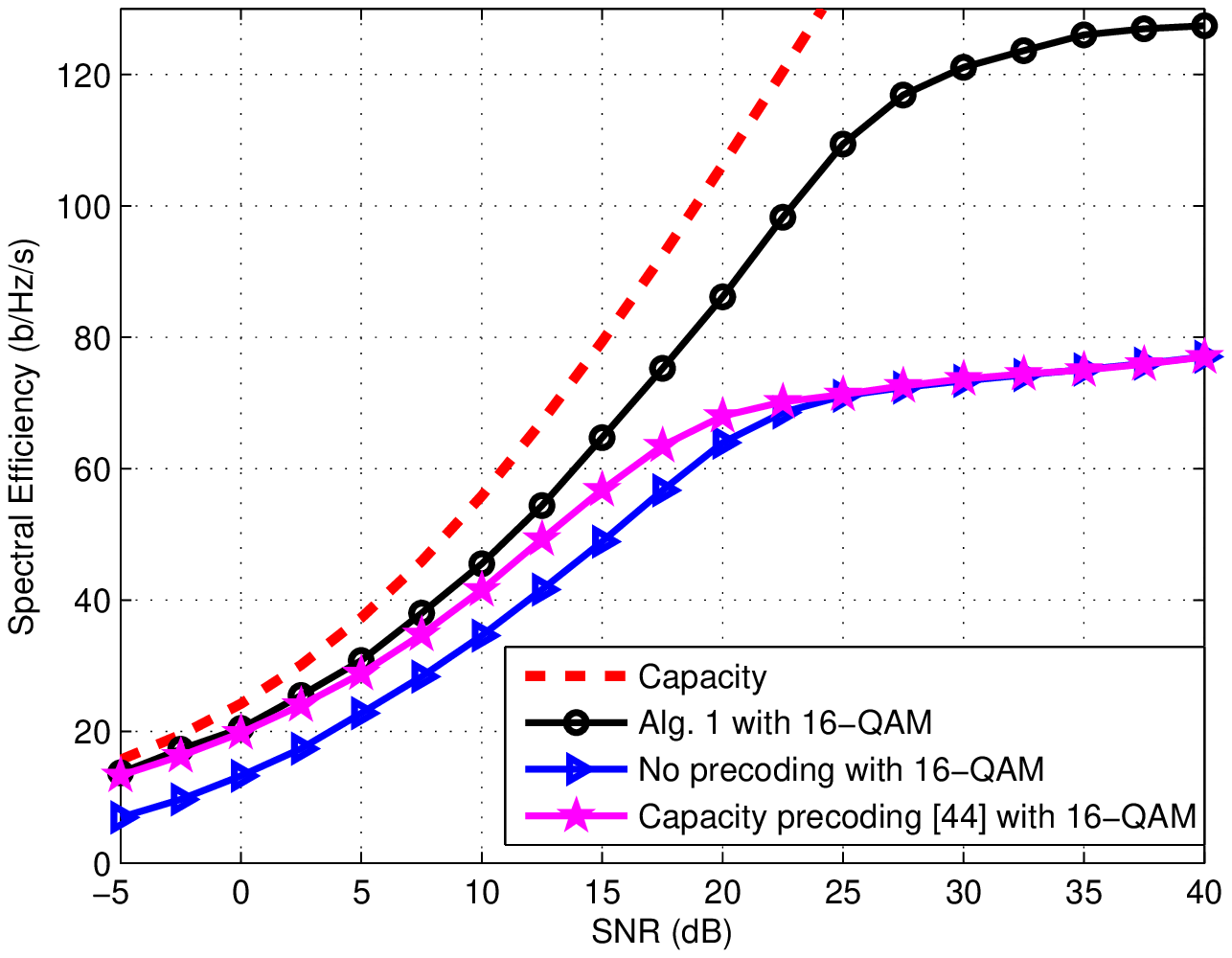}
 \captionstyle{flushleft}
  \vspace{-10pt}
\caption{Spectral efficiency vs. SNR for the 3GPP SCM (urban scenario, half-wavelength antenna spacing, $36$ km/h) for different precoder designs with $N_{\mathrm t} = N_{\mathrm r} =32$, $N_{\mathrm s} = 2$, and 16-QAM.}
\label{16QAM_SCM_32}
\end{figure}

Figs. \ref{QPSK_SCM_32} and \ref{16QAM_SCM_32} present further results for $N_{\mathrm t} = N_{\mathrm r} = 32$ with QPSK and 16-QAM, respectively.
We set $N_{\mathrm s} =4$ for the former and $N_{\mathrm s} = 2$ for the latter.
When $N_{\mathrm t} =32$, the computational complexity of  calculating the
ergodic spectral efficiency in (\ref{eq:Mutual_Info}) scales with
$4^{64}$ and $16^{64}$ for QPSK and 16-QAM, respectively, which is prohibitive.
Algorithm 1, in contrast, can be executed with very satisfactory performance.

\begin{figure}[!t]
\centering
\includegraphics[width=0.8\textwidth]{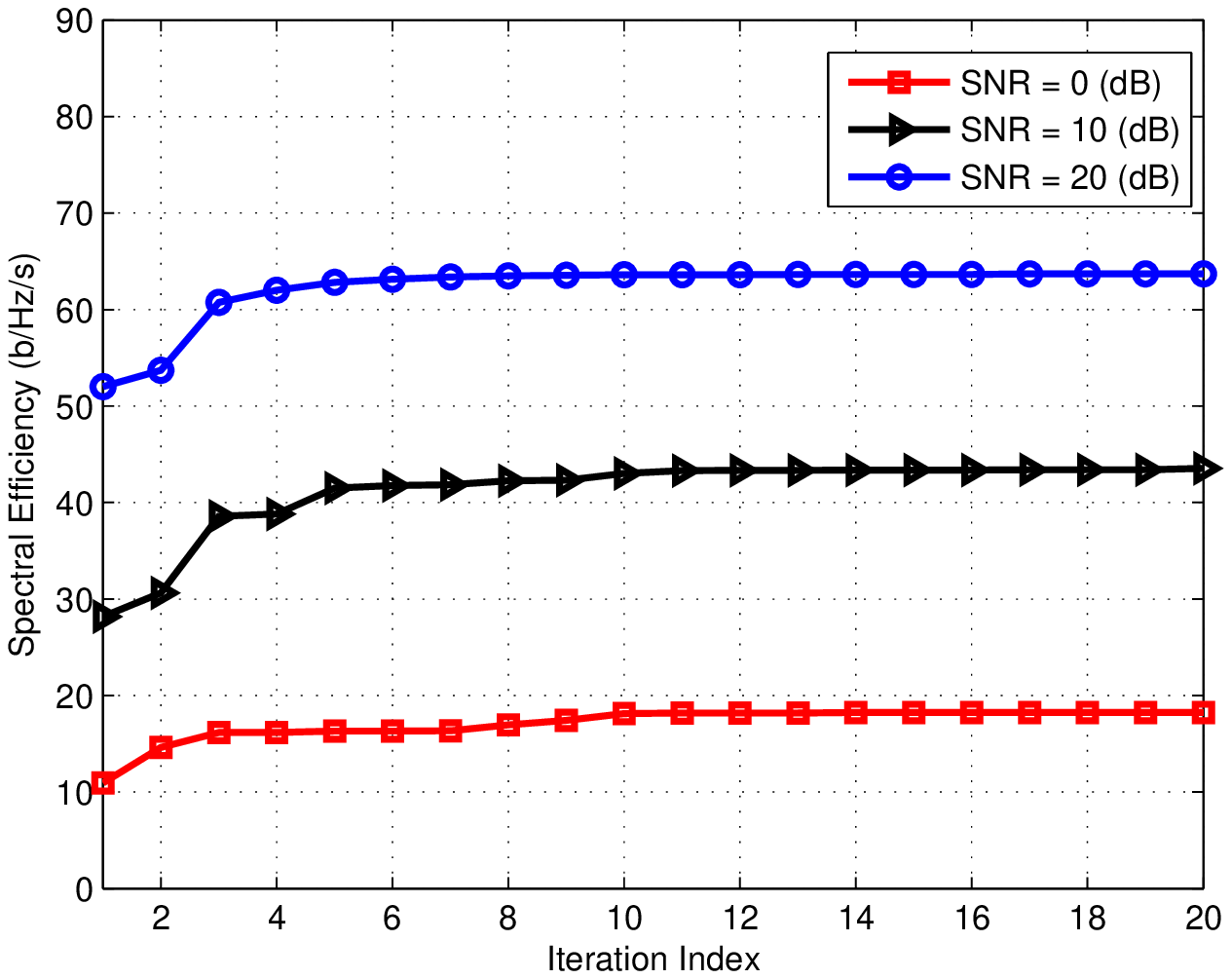}
 \captionstyle{flushleft}
  \vspace{-10pt}
\caption{Convergence of Algorithm \ref{Gradient_Pair} for the 3GPP SCM (urban scenario, half-wavelength antenna spacing, $36$ km/h) for different precoder designs with $N_{\mathrm t} = N_{\mathrm r} =32$, $N_{\mathrm s} = 4$, and QPSK.}
\label{QPSK_SCM_32_converge}
\end{figure}

\begin{figure}[!t]
\centering
\includegraphics[width=0.8\textwidth]{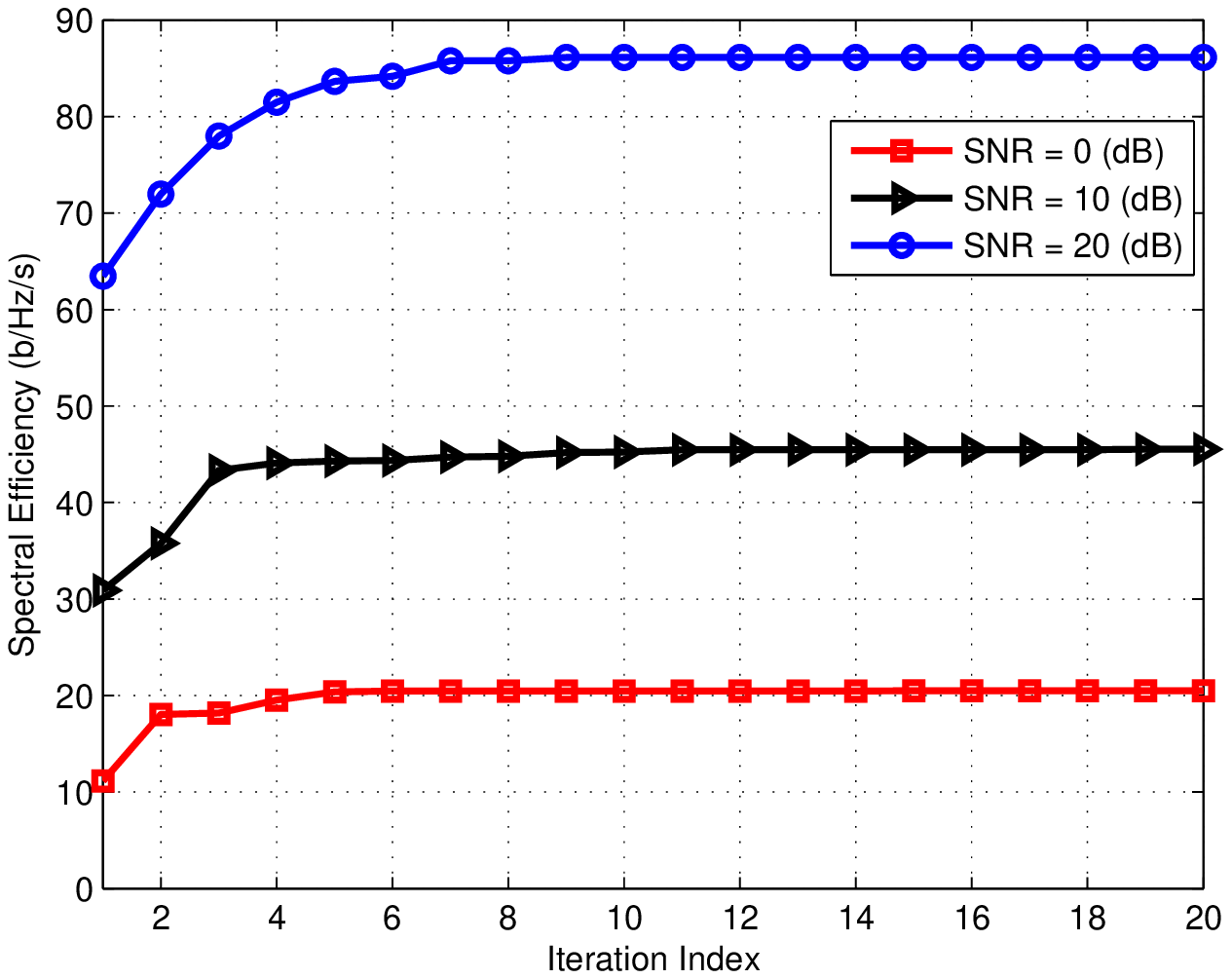}
 \captionstyle{flushleft}
  \vspace{-10pt}
\caption{Convergence of Algorithm \ref{Gradient_Pair} for the 3GPP SCM (urban scenario, half-wavelength antenna spacing, $36$ km/h) for different precoder designs with $N_{\mathrm t} = N_{\mathrm r} =32$, $N_{\mathrm s} = 2$, and 16-QAM.}
\label{16QAM_SCM_32_converge}
\end{figure}

Finally, Figs. \ref{QPSK_SCM_32_converge} and \ref{16QAM_SCM_32_converge} show the convergence of
 Algorithm \ref{Gradient_Pair} at different SNRs for the same settings as in Figs. \ref{QPSK_SCM_32} and \ref{16QAM_SCM_32}, respectively.
In all cases, convergence occurs within 10 iterations.

\section{Conclusion}

With a proper design of $\mathbf{U}_{{\mathrm{\qB}}}$,  $\mathbf{\Lambda}_{{\mathrm{\qB}}}$,
and $\mathbf{V}_{{\mathrm{\bf{B}}}}$,
it is possible to achieve a satisfactory tradeoff between the need to feed into the channel mixings of multiple
finite-cardinality signals and the computational complexity of exploring all possible such mixings.
Building on this idea, an algorithm
has been proposed that---under the 3GPP SCM channel model---exhibits very
good performance with orders-of-magnitude less complexity than complete-search solutions while needing only statistical CSI at the transmitter.
%More refined versions of this algorithm, equipped with alternative subchannel pairing schemes, may perform even better.
%Additional extensions include the applicability to multiuser contexts, or secure communication, as well as other future communication technologies.

The proposed algorithm utilizes the first- and second-order
 channel statistics $\bar\qH$, $\qU_{{\rm R}}$, $\tilde{\qG}$, and $\qU_{{\rm T}}$.
For growing Rice factors, as the channel becomes progressively deterministic, statistical and instantaneous CSI become equivalent; naturally then, the algorithm converges to instantaneous-CSI solutions.
Similarly,
%Algorithm \ref{Gradient_Pair} reduces to precoder designs with instantaneous CSI.
if $\qU_{\rm R} = \qI_{N_{\mathrm r}}$ and $\qU_{\rm T} = \qI_{N_{\mathrm t}}$, then the
algorithm can embrace
precoder designs with estimated CSI, where $\bar\qH$
represents the estimated channel and $\qG$ reflects the power
of the estimation error.
%Also, as indicated in Section \ref{sta:model}, the
%channel model (\ref{eq:Spatial_Cov}) is a very general model.
%Therefore,  Algorithm \ref{Gradient_Pair}
%applies to most of the popular statistical channel models used in communication.
%For particular channel models, the implementation complexity
%of Algorithm \ref{Gradient_Pair} can be further reduced.

%In this work, we have investigated the design of a linear precoder for point-to-point
%MIMO systems for finite alphabet inputs and statistical CSI at the transmitter.
%Based on  an asymptotic expression for the mutual information
%of MIMO systems over a Kronecker fading model in the large antenna regime, we proposed a low complexity iterative
%algorithm to search for the optimal precoder. The computation complexity of the proposed
%algorithm is several orders of magnitude less than that of the conventional design with finite
%alphabet inputs and statistical CSI. Numerical results indicated that the proposed design
%achieves a similar performance  as the design based on the conventional precoder structure for finite alphabet inputs. Also, even for MIMO systems
%with a small number of antennas, the proposed design achieves substantial performance gains
%over existing precoder designs. The performance gain increases as numbers of antennas become large.

\appendices
\section{Proof of Proposition \ref{Proposition_1}}\label{Proof_Proposition_1}
From (\ref{eq:Mutual_Info}), the ergodic mutual information can be expressed as $ I(\qx; \qy)= \calF - N_{\mathrm r}$ with $\calF = {-{E}_{\qy,\qH}\left[\log Z \right. } $ \\
$ {\left.(\qy,\qH)\right]}$ and $Z(\qy,\qH) = { {E}_{\qx}\!\big[e^{- \left\|\qy- \qH \qx \right\|^2}\big] }$. The expectations over $\qy$ and $\qH$ are generally difficult to analyze because of the logarithm. However, these difficulties can be circumvented by rewriting $\calF$ as { \cite[(2.6)]{Nishimori2001}}
\begin{equation}
    \calF = -\lim_{\tau\rightarrow 0}\frac{\partial}{\partial \tau}\log{E}_{\qy,\qH}\left[Z^\tau(\qy,\qH)\right].
\end{equation}
This reformulation allows evaluating ${E}_{\qy,\qH}\left[Z^\tau(\qy,\qH)\right]$ for integer $\tau$, and subsequently for $\tau$ in the vicinity of $0$. This so-called replica method \cite{Edwards1975} has been widely
adopted in statistical physics \cite{Nishimori2001} and
information theory \cite{wen2007asymptotic,Tanaka2002TIT,RMuller2008JSAC,Guo2005TIT,wen2011sum,Mou-03,Zaidel2012TIT,Wu2015TWCOM}.

{ The calculation of $\calF$ via the replica method
consists of the following three steps.
First, we
introduce $\tau$ IID replicated symbols $\qx^{(\alpha)}$, for $\alpha = 0, 1, \ldots,
\tau$, and then, we compute the expectations over $\qy$ and $\qH$ by repeatedly using the Gaussian integral.\footnote{
Let $\mathbf{S} \in \mathbb{C}^{m \times n}$, $\mathbf{A}_1 \in \mathbb{C}^{m \times n}$, and $\mathbf{A}_2 \in \mathbb{C}^{m \times n}$ be complex matrices and $\mathbf{A}_3 \in \mathbb{C}^{n \times n} $ and $\mathbf{A}_4 \in \mathbb{C}^{m \times m} $ positive definite  matrices, respectively. Then, the
following equality holds \cite{Mou-03}, \cite[Lemma 1]{Wu2015TWCOM}:
\begin{equation}
\int  D \mathbf{S}  \, e^{-{\tr}\left(\mathbf{A}_3 \mathbf{S}^H \mathbf{A}_4 \mathbf{S} + \mathbf{A}_1^H \mathbf{S}-\mathbf{S}^H
\mathbf{A}_2\right)} \\
 =\frac{1}{\det(\mathbf{A}_3 \otimes \mathbf{A}_4)}e^{-{\tr}\left(\mathbf{A}_3^{-1} \mathbf{A}_1^H \mathbf{A}_4^{-1} \mathbf{A}_2\right)}.
\end{equation}
}
Second, we simplify the obtained expression for ${E}_{\qy,\qH}\left[Z^\tau(\qy,\qH)\right]$ by assuming that the covariance matrices of the replicas are in symmetry form \cite[Section 2.3]{Nishimori2001}.
Finally, we compute the remaining integrals by using the saddle-point method  (or the method of steepest descent \cite[Section 2.2.4]{Nishimori2001}), and explicitly find the saddle points at $\tau\rightarrow 0$.
In the following, we limit our presentation to the main steps, since analogous
calculations can be found in several earlier works \cite{wen2007asymptotic,Tanaka2002TIT,RMuller2008JSAC,Guo2005TIT,wen2011sum,Mou-03,Zaidel2012TIT,Wu2015TWCOM}.
}

\subsection*{ Step 1 (Replica analysis):}

To compute ${E}_{\qy,\qH} [Z^\tau(\qy,\qH) ]$ it is useful to introduce $\tau$ IID replicated symbols $\qx^{(\alpha)}$, for $\alpha = 0, 1, \ldots,
\tau$, yielding
\begin{equation} \label{eq:ap_sf_E1}
{E}_{\qy,\qH}{\left[Z^{r}(\qy,\qH)\right]}={E}_{\qH,\qX}{\left[\int\!\! D\qy\prod_{\alpha=0}^{\tau}e^{- \left\|\qy- \qH \qx^{(\alpha)}\right\|^2}\right]}
\end{equation}
where ${\qX = \left[ \qx^{(0)} \,\qx^{(1)} \ldots \, \qx^{(\tau)} \right]}$. { The indices $\alpha$ represent different so-called replicas of the system.}
The { integral with respect to} $\qy$ in (\ref{eq:ap_sf_E1}) can be evaluated using the Gaussian integral.
%However, the expectation over $\qH$ and $\qX$ is not easy because $\qH$ are $\qx^{(\alpha)}$ involved.
Then, to disentangle $\qH$ and $\qX$, we introduce a set of random
variables $v_{nm}^{(\alpha)} = \left[\qW\right]_{nm} [\tilde{\qG}]_{nm} \qu_{{\rm T},m}^{H} \qx^{(\alpha)}$. % for $\alpha=0,\dots,\tau$.
Given $[\tilde{\qG}]_{nm}$, $\qu_{{\rm T},m}^{H}$, and $\qx^{(\alpha)}$, it is easily found that the $ v_{nm}^{(\alpha)} $ are Gaussian with zero-mean and covariance $E \big[ v_{nm}^{(\alpha)H} v_{nm}^{(\beta)} \big] = {[\qG ]_{nm} (\qx^{(\alpha)})^H \qT_{m} \qx^{(\beta)} } = Q_{nm}^{(\alpha,\beta)}$, $\forall \alpha, \beta$, where $ \qT_{m} = \qu_{{\rm T},m} \qu_{{\rm T},m}^{H}$. Then, we insert an identity that captures
all combinations of the replicas
\begin{equation}
 1 = \int \prod_{n,m}\prod_{0\leq\alpha\leq\beta}^{r}\delta{\left( [\qG ]_{nm} \qx^{(\alpha)H} \qT_{m}\qx^{(\beta)} - Q_{nm}^{(\alpha,\beta)}\right)}
\end{equation}
into (\ref{eq:ap_sf_E1}). % where $\delta(\cdot)$ denotes Dirac's delta.
Let us define $\qQ_{nm} \in \bbC^{(\tau+1)\times (\tau+1)}$ with ${[\qQ_{nm}]_{\alpha\beta} = Q_{nm}^{(\alpha,\beta)}}$, ${\mathbb Q} = \{\qQ_{nm} \}_{\forall n,m}$, $v_{n}^{(\alpha)} = \sum_{m}v_{nm}^{(\alpha)}$, and $\calV = \{ v_{nm}^{(\alpha)} \}$.
Then, (\ref{eq:ap_sf_E1}) can be written as
\begin{equation} \label{eq:ap_sf_E2}
    {E}_{\qy,\qH}\left[Z^{\tau}(\qy,\qH)\right] =\int e^{\calS({\mathbb Q})}d\mu({\mathbb Q})
\end{equation}
where
\begin{align}
 \calS^{(\tau)}({\mathbb Q})&=\log\!\int\! D{\bf y} {E}_{\calV}{\left[\prod_{\tau}e^{-\left\|{\bf y}- \sum_{n} v_{n}^{(\alpha)} \qu_{{\rm R}_n} - \bar{\qH}\qx^{(\alpha)} \right\|^2}\right]}
 \label{eq:G1} \\
 \mu^{(\tau)}({\mathbb Q}) &= {E}_{\qX}{\left[ \prod_{n,m}\prod_{0\leq\alpha\leq\beta}^{r}\delta\left( \left[\qG\right]_{nm}  \qx^{(\alpha)H} \qT_{m}\qx^{(\beta)} - Q_{nm}^{(\alpha,\beta)} \right) \right]}.
 \label{eq:Measure_mu}
\end{align}
{ The integral in (\ref{eq:ap_sf_E2}) can now be estimated by applying the saddle-point method. Therefore, we are left with the evaluation of
$ \calS^{(\tau)}({\mathbb Q})$ and $\mu^{(\tau)}({\mathbb Q})$ which can be computed by applying the techniques in \cite[Appendix A]{Wu2015TWCOM}. Specifically, the evaluation  of $\mu^{(\tau)}({\mathbb Q})$
is exactly identical to that in \cite[(35)]{Wu2015TWCOM}, whereas for $ \calS^{(\tau)}({\mathbb Q})$, additional manipulations for dealing with $\bar{\qH}\qx^{(\alpha)}$ in (\ref{eq:G1}) are required.
}

Because of the Gaussian nature of $v_{nm}^{(\alpha)}$, we can calculate the expectation over $\calV$ after integrating over $\qy$ in (\ref{eq:G1}). Meanwhile, we apply the inverse Laplace transform of $\delta(\cdot)$\footnote{
The inverse Laplace transform of the $\delta$-function is given by { \cite[(5.140)]{Nishimori2001}}
\begin{equation*}
    \delta(x) = \frac{1}{2\pi j}\int_{-j \infty + t}^{j \infty + t} e^{\tilde{Q} x} d \tilde{Q}, ~\forall t \in \bbR.
\end{equation*}
} to (\ref{eq:Measure_mu}) by
introducing auxiliary variables $\tilde\qQ_{nm}\in {\mathbb C}^{(\tau + 1)\times(\tau + 1)}$ and letting $\tilde{{\mathbb Q}} = \{\tilde\qQ_{nm}\}_{\forall n,m}$. The remaining integrals over $({\mathbb Q},\tilde{{\mathbb Q}}) $ can be
evaluated via the saddle point method yielding
\begin{equation} \label{eq:calF1}
    {\cal F} = -\lim_{\tau\rightarrow 0}\frac{\partial}{\partial \tau} \max_{{\mathbb Q},\tilde{\mathbb Q}}
    \left\{ {\cal F}^{(\tau)} \right\}
\end{equation}
with ${\cal F}^{(\tau)} = \calS^{(\tau)} + \calJ^{(\tau)}$, where
\begin{align}
    \calS^{(\tau)} &= - N_{\mathrm r} \log(\tau+1) - \log\det\left(\qI_{N_{\mathrm r}(\tau + 1)} + \qQ {\bf \Sigma} \otimes \qR \right) \label{eq:defS1}
    \end{align}
\begin{align}
  \calJ^{(\tau)} & = \max_{\tilde{{\mathbb Q}}} \Bigg\{ \sum_{n,m} \tr \left(\tilde\qQ_{nm}\qQ_{nm}\right)
    \notag -\log{E}_{\qX} \left[  e^{\sum_{m}\tr \left( \sum_{n} \left[\qG\right]_{nm} \tilde\qQ_{nm}\qX^H\qT_{m}\qX\right)} \right.  \\
    & \left.\cdot e^{ {\sf vec}(\bar\qV)^H\left[\left(\qQ\otimes\qR\right)^{-1}\left(\left(\qQ{\bf \Sigma}\otimes\qR+\qI_{N_{\mathrm r}(\tau + 1)}\right)^{-1}-\qI_{N_{\mathrm r}(\tau + 1)} \right)\right]{\sf vec}(\bar\qV)  \!} \right]  \!
    \Bigg\} \label{eq:defJ},
\end{align}
${\bf \Sigma}=\qI_{\tau+1}-\frac{1}{(\tau+1)}{\bf 1}{\bf 1}^T$, $\qQ \otimes \qR =\sum_{n}\left(\sum_{m} \qQ_{nm}\right) \otimes \qR_{n}$, $\qR_{n} = \qu_{{\rm
R},n}\qu_{{\rm R},n}^{H}$, and $\bar{\qV} = \bar{\qH} \qX$. For the case with no LOS, the last exponential term in the last line of (\ref{eq:defJ}) disappears
{ \cite[(39)]{Wu2015TWCOM}.} Hence, the LOS makes a nontrivial difference.

\subsection*{ Step 2 (Replica symmetry assumption):}

The extremum over $({\mathbb Q},\tilde{{\mathbb Q}}) $ in (\ref{eq:calF1}) can be obtained
by seeking the point of zero gradient, yielding a set of saddle-point equations. However, explicit expressions for the saddle points are not forthcoming. Therefore, we assume that the saddle points exhibit the replica symmetry (RS) form { \cite[(41) and (42)]{Wu2015TWCOM}} ${\qQ_{nm} =q_{nm}{\bf 11}^T+(c_{nm}-q_{nm})\qI_{\tau+1}}$ and $\tilde\qQ_{nm} =\tilde{q}_{nm}{\bf 11}^T+(\tilde{c}_{nm}-\tilde{q}_{nm})\qI_{\tau+1}$.
Based on RS, $q_{nm}$,$c_{nm}$, $\tilde{q}_{nm}$, and $\tilde{c}_{nm}$ are four parameters that need to be determined.

\subsection*{ Step 3 (Saddle point):}
$q_{nm}$, $c_{nm}$, $\tilde{q}_{nm}$, and $\tilde{c}_{nm}$ can be
obtained by inserting the RS into ${\cal F}^{(\tau)}$ and equating the partial derivatives of the corresponding ${\cal F}^{(\tau)}$ to zero. In this case, it can be verified that $\tilde{c}_{nm} = 0$. Let $\gamma_{nm} = \tilde{q}_{nm}$ and $\psi_{nm} = (c_{nm}-q_{nm})/\left[\qG\right]_{nm}$. Finally, at $\tau=0$, ${\cal F}$ can be expressed as
\begin{align}\label{eq:ap_GenFree}
{\cal F} &\simeq  I\left( \qx;\qz \big| \sqrt{\qXi} \right) +\log\det\left(\qI_{N_{\mathrm r}}+\qR \right)  -\sum_{n,m} \gamma_{nm} \left[\qG\right]_{nm} \psi_{nm} + N_{\mathrm r}
\end{align}
where $\qXi = \qT + \bar{\qH}^H \left(\qI+ \qR\right)^{-1} \bar{\qH}$ with $\qT = \sum_{m} (\sum_{n}\gamma_{nm}) \qT_{m}$, $\qR = \sum_{n} (\sum_{m}\psi_{nm}) \qR_{n}$.
Equating the partial derivatives of ${\cal F}$ over $\gamma_{nm}$ and $\psi_{nm}$, we
obtain $\gamma_{nm} = \gamma_{m}$ and $\psi_{nm} = \psi_{n}$ as given in (\ref{eq:Varsigma_k-MSE}). Note that, since $\gamma_{nm}$ and $\psi_{nm}$ are independent of $m$ and $n$, respectively, we have replaced them with $\gamma_{m}$ and $\psi_{n}$ in (\ref{eq:Varsigma_k-MSE}).
Using $ I(\qx; \qy)= \calF - N_{\mathrm r}$ as given at the beginning of this appendix, we finally obtain Proposition \ref{Proposition_1}.

\section{Proof of Proposition \ref{prop:opt_structure}}\label{Proof_prop:opt_structure}
Consider the optimization problem
\begin{equation} \label{eq:max_problem}
\begin{aligned}
 \mathop {\max }\limits_{{\bf{B}}} \ & \hspace{0.5cm} I_{\rm asy}\left( {{\bf{x}};{\bf{y}} } \right)  \\
 \text{s.t.} \hspace{0.25cm} & \hspace{0.5cm} \text{tr} \left( {\mathbf{B} \mathbf{B}^H}\right) \leq P .
\end{aligned}
\end{equation}

The equivalent channel matrix $\qXi$ in (\ref{eq:eqChxi}) is a
function of the precoder $\mathbf{B}$ through the coupled equations (\ref{eq:eqChMatrixTR})--(\ref{eq:Varsigma_k-MSE}).
Thus, the derivation in \cite[App. A]{zeng2012linear} that requires the channel matrix to be independent of
the precoder cannot be applied directly here. %  to prove Proposition \ref{prop:opt_structure}.

To solve (\ref{eq:max_problem}), we establish the Lagrangian  function for (\ref{eq:max_problem})
in terms of the
precoder $\qB$ as
\begin{equation}\label{eq:cost_function_mac}
g\left( {{\bf{B}}} \right) =  - I_{\rm asy}\left( {{\bf{x}};{\bf{y}} } \right)  +  \kappa \left[ {{\rm{tr}}\left( {{\bf{B}} {\bf{B}}^H } \right) - P } \right]
\end{equation}
where  $\kappa$ is a Lagrange multiplier.

The Karush-Kuhn-Tucker condition \cite{Boyd2004} dictates that $\nabla_{\mathbf{B}}g\left( {{\bf{B}}} \right) = \mathbf{0}$ or, equivalently, that
\begin{equation}\label{eq:cost_function_mac_nec}
  - \nabla_{\mathbf{B}} I_{\rm asy}\left( {{\bf{x}};{\bf{y}} } \right)  +  \kappa \qB = \mathbf{0}.
\end{equation}
For ease of exposition, we define $I_1(\qB) =  I\left( {{\bf{x}};{\bf{z}}} \right)$.
For $I_{\rm asy}\left( {{\bf{x}};{\bf{y}} } \right)$ in (\ref{eq:GAUMutuall_2}), the parameters affected by
the perturbation of $\qB$ are $\left\{I_1(\qB), \gamma_m, \psi_n \right\}$.
According to the chain rule, the gradient of $I_{\rm asy}\left( {{\bf{x}};{\bf{y}} } \right)$ with respect
to $\qB$ is given by
\begin{multline}\label{eq:grad_Iasy}
 \nabla _{{\bf{B}}} I_{{\rm{asy}}} \left( {{\bf{x}};{\bf{y}}} \right) = {\frac{{\partial I_{{\rm{asy}}} \left( {{\bf{x}};{\bf{y}}} \right)}}{{\partial I_1(\qB)}}} \nabla _{{\bf{B}}} I_1(\qB)   + \log _2 e  \left( {  \sum\limits_{m = 1}^{N_{\mathrm r}}{\frac{{\partial I_{{\rm{asy}}} \left( {{\bf{x}};{\bf{y}}} \right)}}{{\partial \gamma_m} }  \nabla _{{\bf{B}}} \gamma_m}  +  \sum\limits_{n = 1}^{N_{\mathrm t}} {\frac{{\partial I_{{\rm{asy}}} \left( {{\bf{x}};{\bf{y}}} \right)}}{{\partial \psi_n }}\nabla _{{\bf{B}}} \psi_n } }\right).
\end{multline}

The relationship between the channel $\qXi$ in (\ref{eq:eqChxi}) and
the precoder $\qB$ is determined by the parameters $\gamma_m$ and $\psi_n$ in (\ref{eq:Varsigma_k-MSE}).
Hence, when calculating the first term on the right-side of (\ref{eq:grad_Iasy}), $\qXi$
is regarded as independent of $\qB$.  Also, from the definitions of $\gamma_m$ and $\psi_n$
in Appendix \ref{Proof_Proposition_1}, we have that
\begin{align}\label{eq:grad_I_gamma_psi}
 \frac{{\partial I_{{\rm{asy}}} \left( {{\bf{x}};{\bf{y}}} \right)}}{{\partial \gamma_m }} = 0, \  \frac{{\partial I_{{\rm{asy}}} \left( {{\bf{x}};{\bf{y}}} \right)}}{{\partial \psi_n }} =0.
\end{align}
 As a result, based on (\ref{eq:cost_function_mac_nec})--(\ref{eq:grad_I_gamma_psi}) and \cite[(22)]{Palomar2006TIT}, the optimal precoder should satisfy
the condition
\begin{equation}\label{eq:cost_function_mac_nec_2}
  \kappa \qB =  \qXi \qB  \boldsymbol{\Omega}.
\end{equation}
Using the eigenvalue decomposition $\boldsymbol{\Omega} = \qU_{\boldsymbol{\Omega}} \qLambda_{\boldsymbol{\Omega}} \qU_{\boldsymbol{\Omega}}^H$, we can rewrite
(\ref{eq:cost_function_mac_nec_2}) as
\begin{equation}\label{eq:cost_function_mac_nec_21}
  \kappa  \qU_{\qXi}^H \qB \qU_{\boldsymbol{\Omega}} =  \qLambda_{\mathrm \qXi} \qU_{\qXi}^H  \qB \qU_{\boldsymbol{\Omega}} \qLambda_{\boldsymbol{\Omega}}.
\end{equation}
Define  $\qQ = \qU_{\qXi}^H \qB \qU_{\boldsymbol{\Omega}}$. Then, we have
\begin{equation}\label{eq:cost_function_mac_nec_22}
  \kappa  \qQ =  {\rm diag}\left({\qLambda_{\qXi}}\right) {\rm diag}\left({\qLambda_{\mathrm \Omega}}\right)^T \odot \qQ
\end{equation}
which is equivalent to
\begin{equation}\label{eq:cost_function_mac_nec_3}
  \kappa  \left[\qQ \right]_{mn} =  \left[{\qLambda_{\qXi}} \right]_{mm} \left[{\qLambda_{\boldsymbol{\Omega}}} \right]_{nn} \left[\qQ \right]_{mn}.
\end{equation}
The eigenvalues of $\qXi$ and $\qOmega$ are distinct with probability one. Therefore,  the equality
$\kappa = \left[{\qLambda_{\qXi}} \right]_{mm} \left[{\qLambda_{\boldsymbol{\Omega}}} \right]_{nn}$ can be satisfied
for at most $N_{\mathrm t}$ pairs of $(m,n)$, each corresponding to different $m$ and $n$.
 For other pairs of $(m,n)$, $\left[\qQ \right]_{mn} = 0$  so that
(\ref{eq:cost_function_mac_nec_3}) can hold. As a result, $\qQ $ has
 at most one nonzero entry in each row and in each column. Thus, $\qQ$ can
 be written as
 \begin{equation}\label{eq:cost_function_mac_nec_4}
 \qQ =  \qLambda \qPi
  \end{equation}
 where $\qLambda$ is diagonal and $\qPi$ is a permutation matrix. Recalling the definition of
 $\qQ $,  the optimal precoder is $\qB = \qU_{\qXi}  \qLambda \qPi \qU_{\boldsymbol{\Omega}}^H$.
 %The proof is now completed.

%\bibliographystyle{IEEETran}
%\bibliography{IEEEabrv,Reference}

%\begin{figure}[!ht]
%\centering
%\includegraphics[width = 0.8\textwidth]{figure/bc_{\mathrm r}x}
%\end{figure}
\end{document}